\documentclass[onecolumn]{revtex4-1}

\usepackage{float}
\usepackage{amsmath,amssymb}
\usepackage{graphics,graphicx,color,colortbl}
\usepackage{booktabs}
\usepackage{pdfpages}
\newcommand{\red}{\textcolor{black}}

\usepackage{ulem}

\begin{document}

\title{Cell assembly dynamics of sparsely-connected inhibitory networks:\\
a simple model for the collective activity of striatal projection neurons}

\author{David Angulo-Garcia}
\email{david.angulo@fi.isc.cnr.it}
\affiliation{CNR - Consiglio Nazionale delle Ricerche -
Istituto dei Sistemi Complessi, via Madonna del Piano 10,
I-50019 Sesto Fiorentino, Italy}
\author{Joshua D. Berke}
\email{jdberke@umich.edu}
\affiliation{Department of Psychology, 
University of Michigan, Ann Arbor,
530 Church St., Ann Arbor, MI 48104, USA}
\author{Alessandro Torcini}%
\email{alessandro.torcini@cnr.it}
\affiliation{Aix-Marseille Universit\'{e}, Inserm, INMED UMR 901 and Institut de Neurosciences des Syst\`{e}mes UMR 1106, 13000 Marseille, France}
\affiliation{Aix-Marseille Universit\'{e}, Universit\'{e} de Toulon, CNRS, CPT, UMR 7332, 13288 Marseille, France}
\affiliation{CNR - Consiglio Nazionale delle Ricerche -
Istituto dei Sistemi Complessi, via Madonna del Piano 10,
I-50019 Sesto Fiorentino, Italy}
\affiliation{INFN Sez. Firenze, via Sansone, 1 - I-50019 Sesto Fiorentino, Italy}

\begin{abstract}
Striatal projection neurons form a sparsely-connected inhibitory network, and this arrangement may be essential for the appropriate temporal organization of behavior. Here we show that a \red{simplified}, sparse inhibitory network of Leaky-Integrate-and-Fire neurons can reproduce \red{some} key features of striatal population activity, as observed in brain slices [{\it Carrillo-Reid et al., J. Neurophysiology {\bf 99} (2008) 1435--1450}]. In particular we develop a new metric to determine the conditions under which sparse inhibitory networks form anti-correlated cell assemblies with \red{ time-varying activity} of individual cells. \red{We find that under these conditions the network displays an input-specific sequence of cell assembly switching, that effectively discriminates similar inputs. Our results support the proposal [{\it Ponzi and Wickens, PLoS Comp Biol {\bf 9} (2013) e1002954}] that GABAergic connections between striatal projection neurons allow stimulus-selective, temporally-extended sequential activation of cell assemblies. Furthermore, we help to show how altered intrastriatal GABAergic signaling may produce aberrant network-level information processing in disorders such as Parkinson's and Huntington's diseases.}
\\
\\
\\
\\
\\
\\
\textbf{Author Summary}\\
\red{
Neuronal networks that are loosely coupled by inhibitory connections can exhibit potentially useful properties. These include the ability to produce slowly-changing activity patterns, that could be important for organizing actions and thoughts over time. The striatum is a major brain structure that is critical for appropriately timing behavior to receive rewards. Striatal projection neurons have loose inhibitory interconnections, and here we show that even a highly simplified model of this striatal network is capable of producing slowly-changing activity sequences. We examine some key parameters important for producing these dynamics, and help explain how changes in striatal connectivity may contribute to serious human disorders including Parkinson's and Huntington's diseases.}
\end{abstract}

\maketitle

\section{Introduction}

The basal ganglia are critical brain structures for behavioral control, whose organization has been highly conserved 
during vertebrate evolution~\cite{stephenson2011}. \red{Altered activity} of the basal ganglia underlies a wide range 
of human neurological and psychiatric disorders, but the specific computations normally performed by these circuits remain elusive. 
The largest component of the basal ganglia is the striatum, which appears to have a key role in adaptive decision-making based 
on reinforcement history~\cite{daw2006}, and in \red{behavioral timing on scales from tenths of seconds to tens of seconds~\cite{merchant2013}.}

The great majority ($>90\%$) of striatal neurons are GABAergic medium spiny neurons (MSNs), which project to other basal ganglia structures but also make local collateral connections within striatum~\cite{west1996estimationMSN,oorschot1996total}. 
These local connections were proposed in early theories to achieve action selection through strong winner-take-all lateral inhibition ~\cite{groves1983,beiser1998}, but this idea fell out of favor once it became clear that MSN connections are actually sparse (nearby connection probabilities $\simeq 10-25 \%$~\cite{tunstall2002inhibitory,taverna2004sparseStriatum}), unidirectional and relatively weak~\cite{tepper2004gabaergic,jaeger1994surround}. \red{
Nonetheless, striatal networks are intrinsically capable of generating 
sequential patterns of cell activation, even in brain slice preparations 
without time-varying external inputs~\cite{carrillo2008encoding,carrillo2009motifs}.}
\red{Following  previous experimental evidence that collateral inhibition can help organize MSN firing~\cite{Guzman2003MSN},
an important recent set of modeling studies argued that the sparse connections between MSNs, though individually weak, can collectively mediate sequential switching between cell assemblies~\cite{ponzi2010sequentially,ponzi2012input}.}
\red{It was further hypothesized that these connections
may even be optimally configured for this purpose~\cite{ponzi2013optimal}.} 
This proposal is of high potential significance, since sequential dynamics may be central to the striatum's functional role in the organization and timing of behavioral output~\cite{berke2009,carrillo2011}.
  
In their work~\cite{ponzi2010sequentially,ponzi2012input,ponzi2013optimal}, Ponzi and Wickens used conductance-based model neurons (with persistent $Na^+$ and $K^+$ currents~\cite{izhikevich2008dynamical}), in proximity to a bifurcation from a stable fixed point to a tonic firing regime. We show here that networks based on simpler leaky integrate-and-fire (LIF) neurons can also exhibit sequences of cell assembly activation. \red{This simpler model, together with a novel measure of structured bursting,} allows us to more clearly identify the \red{critical parameters} needed to observe dynamics resembling that of the striatal MSN network.
Among other results, we show that the duration of GABAergic post-synaptic currents is crucial 
for the network$^\prime$s ability to discriminate different input patterns.
\red{A reduction of the post-synaptic time scale, analogous to that observed for IPSCs
of MSNs in mouse models of Huntington's disease (HD)~\cite{cummings2010},
leads in our model to alteration of single neuron and population dynamics 
typical of striatal dynamics in symptomatic HD mice~\citep{miller2008dysregulated}.} 
   Finally, we qualitatively replicate the observed response of striatal networks in brain slices to altered excitatory drive \red{and to reduction of GABAergic transmission between axon collaterals of striatal neurons~\cite{carrillo2008encoding}. The latter effect can be induced by dopamine loss~\cite{lopez2013}, therefore our results may help generate new insights into the aberrant 
activity patterns observed in Parkinson's disease (PD).}

\begin{figure*}
\centering
\includegraphics[width=0.35\textwidth]{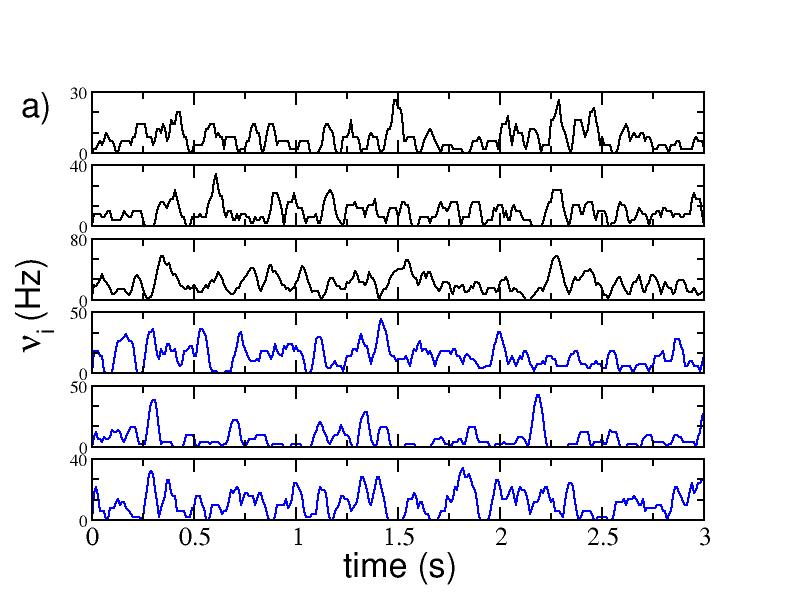}
\includegraphics[width=0.32\textwidth]{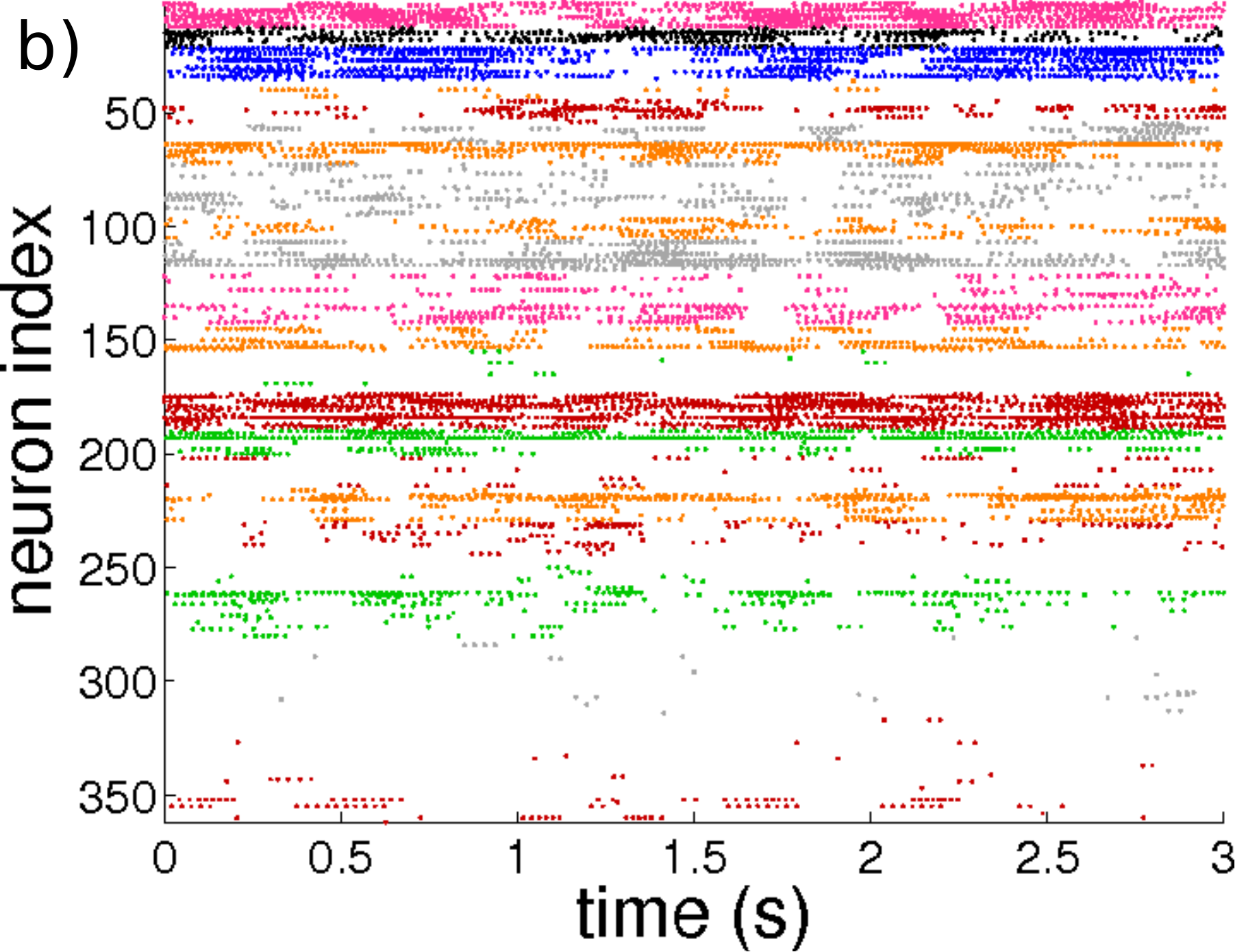}
\includegraphics[width=0.29\textwidth]{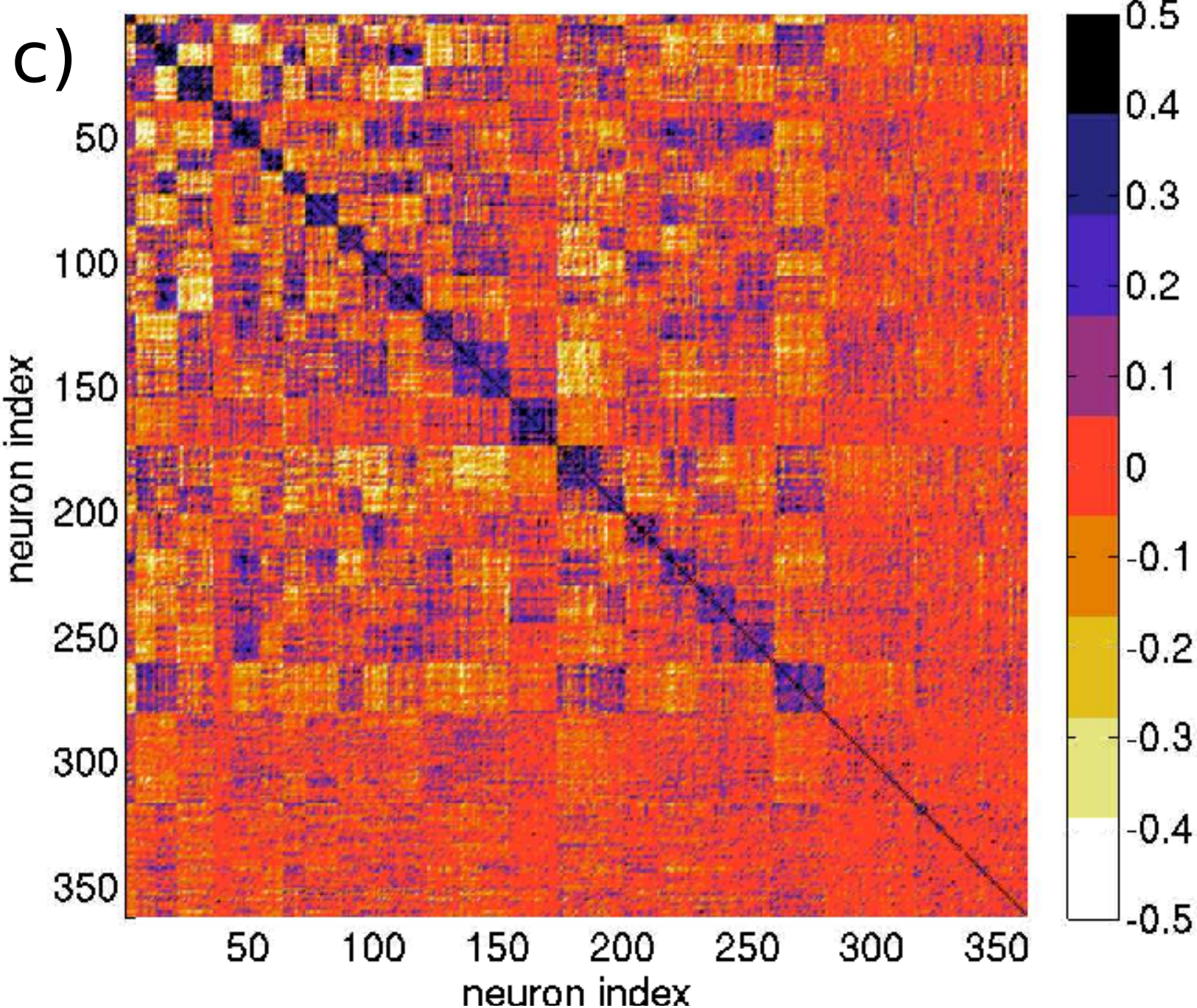}\\
\caption{{\bf Cell activity characterization.} \red{a) Firing rates $\nu_i$ 
of 6 selected neurons belonging to two anti-correlated assemblies, 
the color identifies the assembly and the colors correspond to the one used 
in b) for the different clusters;}
b) raster plot activity, the firing times are colored 
according to the assembly the neurons belong to; c) cross-correlation matrix 
$C(\nu_i,\nu_j)$ of the firing rates. \red{The neurons in panel b) and c)
are clustered according to the correlation of their firing rates by employing the \textit{k-means} algorithm; the clusters are ordered in terms of their average correlation (inside each cluster)
from the highest to the lowest one (for more details see Methods).} The firing rates are calculated over overlapping time windows of duration 1 s, 
the origins of successive windows are shifted by 50 ms. 
The system is evolved during $10^7$ spikes, after discarding an initial transient of $10^5$ spike events. Other parameters used in the simulation: $g = 8$, $K = 20$, $N = 400$, $k_{mean} = N_{act}/15$, $\Delta V = 5$ mV and $\tau_{\alpha} = 20$ ms. \red{The number of active neurons is
370, corresponding to $n^* \simeq 93$ \%.}}
\label{fig:PonziBenchmark}
\end{figure*}

\section{Results} 

\subsection*{Measuring cell assembly dynamics}

The model is composed of $N$ leaky integrate-and-fire (LIF) inhibitory neurons~\cite{burkitt2006LIFreviewI,burkitt2006LIFreviewII},
with each neuron receiving inputs from a randomly selected 5$\%$ of the other neurons (i.e. a directed Erd\"os-Renyi graph with constant in-degree $K = pN$, where $p = 5\%$)~\cite{Brunel2000Sparse}. \red{The inhibitory post-synaptic potentials (PSPs) are schematized as $\alpha$-functions characterized by a decay time $\tau_\alpha$ and a peak amplitude $A_{PSP}$.}
In addition, each neuron $i$ is subject to an excitatory input current mimicking the cortical and thalamic inputs received by the striatal network. \red{In order to obtain firing periods
of any duration the excitatory drives are tuned to drive the
neurons in proximity of the supercritical bifurcation 
between the quiescent and the firing state, similarly to~\cite{ponzi2010sequentially}}.
\red{Furthermore, our model is integrated exactly between a spike emission and the successive one by rewriting the time evolution of the network as an event-driven map~\cite{Zillmer2007} (for more details see Methods).}

\red{Since we will compare most of our findings with the results reported in a previous
series of papers by Ponzi and Wickens (PW)~\cite{ponzi2010sequentially,ponzi2012input,ponzi2013optimal}
it is important to stress the similarities and differences between the two models. The model
employed by PW is a two dimensional conductance-based model with a potassium and a sodium channel~\cite{izhikevich2008dynamical}, our model is simply a current based LIF model.
The parameters of the PW model are chosen so that the
cell is in proximity of a saddle-node on invariant circle (SNIC) bifurcation 
to guarantee a smooth increase of the firing period when passing from the
quiescent to the supra-threshold regime, without a sudden jump in the firing rate.
Similarly, in our simulations the parameters of the LIF model are chosen in proximity of 
the bifurcation from silent regime to tonic firing. In the PW model the PSPs are assumed to
be exponentially decaying, in our case we considered $\alpha$-functions. 
}

In particular, we are interested in \red{selecting} model parameters for which uniformly distributed inputs $I = \{ I_i \}$, where $I_i \in [V_{th}, \, V_{th} + \Delta V]$, produce cell assembly-like sequential patterns in the network. The main aspects of the desired activity can be summarized as follows: (i) single neurons should exhibit large variability in firing rates ($CV > 1$);
(ii) the dynamical evolution of neuronal assemblies should reveal strong correlation within an assembly and anti-correlation with neurons out of the 
assembly. As suggested by many authors~\cite{tepper2004gabaergic,plenz2003} the dynamics of MSNs cannot be explained in terms of a {\it winners take all} (WTA) mechanism, which would imply a \red{small} number of highly firing neurons, while the remaining would be silenced. \red{Therefore we will search, in addition to the requirements (i) and (ii), for a regime where a substantial fraction of neurons are actively involved in the network dynamics. This represents a third criterion (iii) to be 
fulfilled to obtain a striatum-like dynamical evolution.}

Fig.~\ref{fig:PonziBenchmark} shows an example of such dynamics for the LIF model, with three pertinent similarities to previously observed dynamics of real MSN networks~\cite{carrillo2008encoding}. Firstly, cells are organized into correlated groups, and these groups are mutually anticorrelated (as evident from the cross-correlation matrix of the firing rates reported in Fig.~\ref{fig:PonziBenchmark} (c)).
\red{Secondly, individual cells within groups show irregular firing as shown in Fig.~\ref{fig:PonziBenchmark} (a). This aspect is reflected in a coefficient of variation ($CV$) of the inter-spike-intervals (ISIs) definitely greater than one (see the black curve in Fig.~\ref{fig:CV1_CV2} (b)) as observed experimentally for the dynamics of rat striatum \textit{in-vitro}~\cite{tunstall2002inhibitory,tepper2004gabaergic}. }
\red{Thirdly, the raster plot reported in Fig.~\ref{fig:PonziBenchmark} (b) reveals that a \red{large} fraction of neurons (namely,$\simeq 93$ \%) is actively involved in the neural dynamics.}

\subsection*{A novel metric for the structured cell assembly activity}

\red{
The properties (i),(ii), and (iii), characterizing MSN activity, can be quantified in terms of a single scalar metric $Q_0$, as follows:}
\begin{equation}
\label{eq:Q0Indicator}
Q_0 \equiv \langle CV \rangle_N \times \sigma(C(\nu_i,\nu_j)) \times n^* 
\quad ;
\end{equation}
where  $\langle \cdot \rangle _ N$ denotes average over the ensemble of $N$ neurons, $n^* = N_{act}/N$ is the fraction of active neurons $N_{act}$ out of the total number, $C(\nu_i,\nu_j)$ is the $N\times N$ zero-lag cross-correlation matrix between all the pairs of 
single neuron firing rates $(\nu_i,\nu_j)$, and $\sigma (\cdot)$ is the standard deviation
of this matrix (for details see Methods). \red{We expect that good parameter values for our model
can be selected by maximizing $Q_0$.} 

\red{Our metric is inspired by a metric introduced to identify the level of cluster synchronization and organization for a detailed  striatal microcircuit model~\cite{humphries2009dopamine}.
However, that metric is based on the similarity among the point-process
spike trains emitted by the various neurons, whereas $Q_0$ uses correlations between 
firing rate time-courses. Furthermore, $Q_0$ takes also in account the variability of the firing rates, by including the average $CV$ in Eq. \eqref{eq:Q0Indicator}, an aspect of the MSN dynamics 
omitted by the metric employed in \cite{humphries2009dopamine}.} \red{Within biologically meaningful ranges, we find values of the parameters controlling lateral inhibition (namely, the synaptic strength $g$ and the the post-synaptic potential duration $\tau_{\alpha}$) that maximize $Q_0$.
As we show below, the chosen parameters not only produce MSN-like network dynamics but also 
optimize the network$^\prime$s computational capabilities,} in the sense of producing 
a consistent, temporally-structured response, to a given pattern of inputs while  
\red{discriminating between inputs which differ only for a few elements.}

\subsection*{The role of lateral inhibition}

\begin{figure}
\centering
\includegraphics[width=0.47\textwidth]{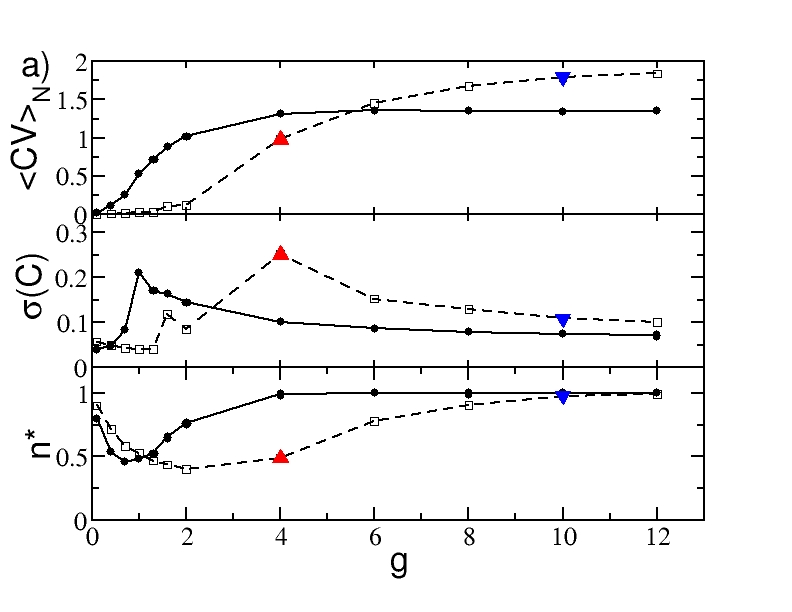}
\includegraphics[width=0.47\textwidth]{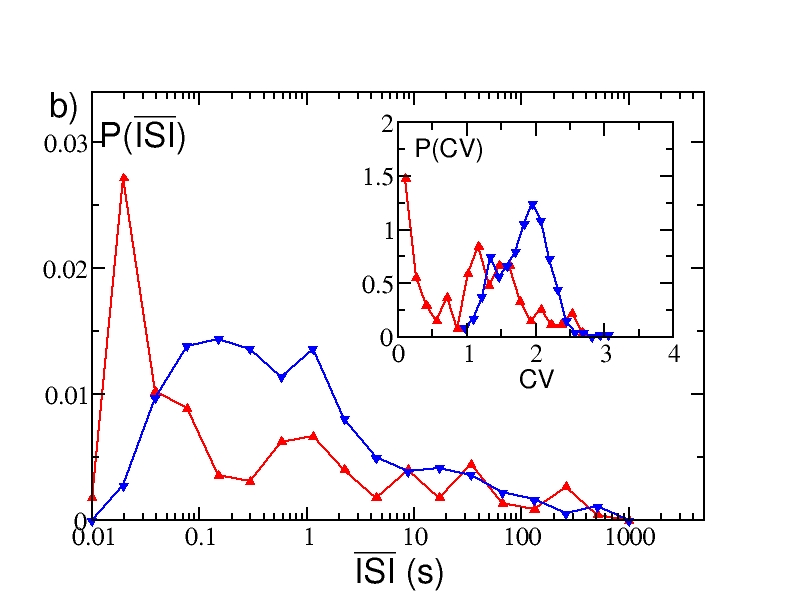}
\includegraphics[width=0.47\textwidth]{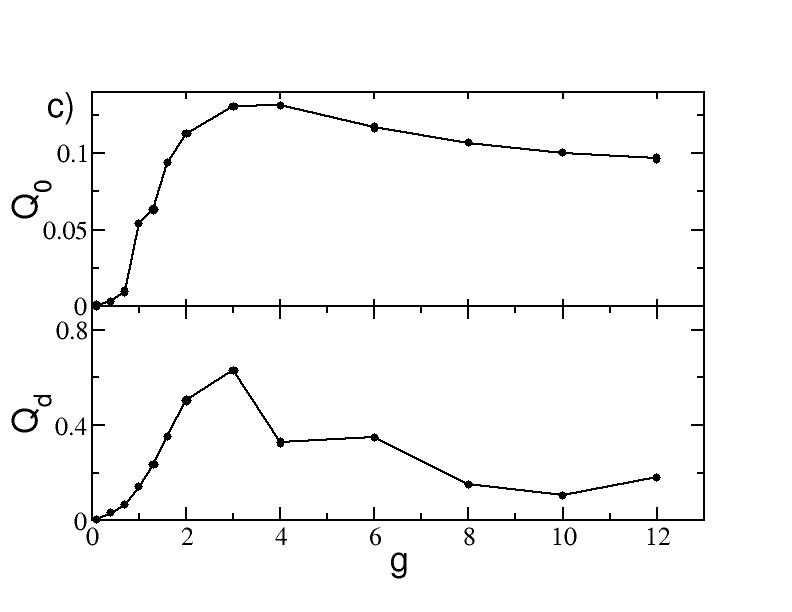}
\includegraphics[width=0.47\textwidth]{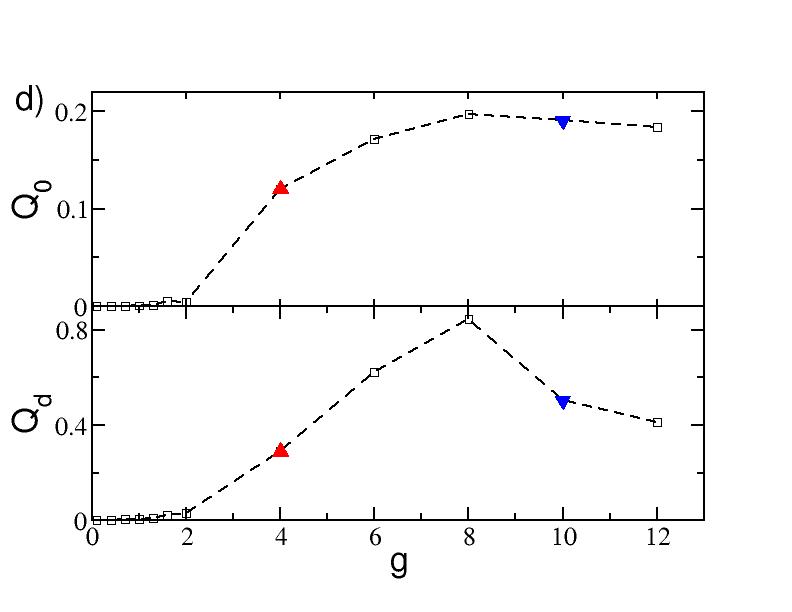}
\vspace{-4mm}
\caption{\textbf{Metrics of structured activity vs lateral inhibition strength.} \red{a) 
Metrics entering in the definition of $Q_0$ versus the synaptic strength $g$. From top to bottom: Averaged coefficient of variation $\langle CV \rangle_N$, standard deviation of the cross-correlation matrix $\sigma(C)$,}
and the fraction of active neurons $n^*$. The solid (dashed) line refers to the case 
$\Delta V = 1$ mV ($\Delta V= 5$ mV). The minimum number of active neurons is
achieved at $g=g_{min}$, this corresponds to a peak amplitude of the PSP $A_{PSP} = 0.064$ mV 
($A_{PSP} = 0.184$ mV) for $\Delta V = 1$ mV ($\Delta V = 5$  mV) (for more details see Methods).
b) Distributions $P(\overline{ISI})$ of the average ISI for a 
fixed $\Delta V= 5$ mV and for two different coupling strengths, $g=4$ (red triangle-up symbol) and $g=10$ (blue triangle-down). Inset, the distribution $P(CV)$ of the $CV$ of the single neurons for the same two cases. c) $Q_0$ and $Q_d$,  as defined in Eqs. \eqref{eq:Q0Indicator} and \eqref{eq:deltaM2}, versus $g$ for $\Delta V = 1$ mV. d) Same as c) for $\Delta V= 5$ mV.
\red{Other parameters as in Fig. \ref{fig:PonziBenchmark}}}
\label{fig:Q0andCV}
\end{figure}

In this sub-section we examine how network dynamics is affected by the strength of inhibitory connections (Fig.~\ref{fig:Q0andCV}). When these lateral connections are very weak (parameter $g$ close to zero), the dominant input to each neuron is the constant excitation. As a result, most individual neurons are active (fraction of active neurons, $n^*$, is close to 1) and firing regularly ($CV$ close to zero). \red{As lateral inhibition is made stronger, some neurons begin to slow down or even 
stop firing, and $n^*$ declines towards a minimum fraction of $\simeq 50\%$ (at $g = g_{min}$)}. \red{As noted by Ponzi and Wickens~\cite{ponzi2013optimal}, this is due to a winner-take-all (WTA) mechanism :} faster-firing neurons depress or even silence the neurons to which they are connected. 
\red{This is evident from the distribution $P(\overline{ISI})$ of the average interspike intervals ($\overline{ISI}$), which is peaked at low firing periods, and from the distribution of the $CV$ 
exhibiting a single large peak at $CV \simeq 0$  (as shown in the insets of Figs.~S2 (a,b) and (d,e)).}

\red{As soon as $g > g_{min}$, the neuronal activity is no longer due only to the
{\it winners}, but also the {\it losers} begin to play a role. 
The {\it winners} are characterized by an effective input $W_i$ which is on average supra-threshold,
while their firing activity is driven by the mean current: {\it winners} are {\it mean-driven}~\cite{renart2007}. On the other hand, {\it losers} are on average below-threshold, and their firing is due to current fluctuations: {\it losers} are  {\it fluctuation-driven}~\cite{renart2007}. 
For more details see Figs.~S2 (c) and (f)).}
\red{This is reflected in the corresponding distribution $P(\overline{ISI})$ (Fig.~\ref{fig:Q0andCV}(b), red curve). The {\it winners} have very short $\overline{ISI}$s (i.e. high firing rates), while the {\it losers} are responsible for the long tail of the distribution extending up to $\overline{ISI} \simeq 10^3$ s. \red{In the distribution of the coefficients of variation (Fig. ~\ref{fig:Q0andCV}(b) inset, red curve) the {\it winners} generate a peak of very low $CV$ (i.e. highly-regular firing),} suggesting that they are not strongly influenced by the other neurons in the network and therefore have an effective input on average supra-threshold. By contrast the {\it losers} are associated with a smaller peak at $CV \simeq 1$, confirming that their firing is due to large fluctuations in the currents.}

Counterintuitively however, further increases in lateral inhibition strength 
result in increased neuronal participation, with $n^*$ progressively returning towards $\simeq 1$.
The same effect was previously reported by Ponzi and Wickens~\cite{ponzi2013optimal} for a different, 
more complex, model. When the number of active neurons returns almost to
100\%, i.e. for sufficiently large coupling $g > g_{min}$, most of the neurons appear to be
below threshold, as revealed by the distribution of the effective inputs $W_i$
reported in Figs.~S2 (c) and (f). Therefore in this case the network firing is essentially 
fluctuation-driven, as a matter of fact the $P(\overline{ISI})$ distribution is now characterized
by a broader distribution and by the absence of the peak at short $\overline{ISI}$
(as shown in Fig.~\ref{fig:Q0andCV} (b), blue line; see also Figs. S2(a) and (d)).
Furthermore the single neuron dynamics is definitely \red{bursting}, as shown
by the fact that the $CV$ distribution  is now
centered around $CV \simeq 2$ (inset of  Fig.~\ref{fig:Q0andCV} (b), blue line; see also
Figs. S2(b) and (e)).

The transition between the two dynamical regimes, occurring at $g = g_{min}$,  
is due to a passage from a state \red{where some {\it winner} neuron}
were  mean-driven and were able to depress all the other neurons, to a state at $g >> g_{min}$
where almost all neurons are fluctuation-driven and contribute to the
network activity. The transition occurs because at $g < g_{min}$ the fluctuations
in the effective input currents $W_i$ are small and insufficient to drive the {\it losers}
towards the firing threshold (as shown in the insets of Fig.~S2 (c) and (f)).
\red{At $g \simeq g_{min}$ the amplitude of the fluctuations becomes sufficient for
some {\it losers} to cross the firing threshold and contribute to the number
of active neurons.} This will also reduce the {\it winners$^\prime$} activity.
\red{For $g >> g_{min}$ the fluctuations of $W_i$ are sufficient to lead almost 
all {\it losers} to fire and no clear distinction between {\it losers}
and {\it winners} remains.} The transition is due to the fact that not only the
average inhibitory action is proportional to the synaptic strength, but also the
amplitude of the current fluctuations increases linearly with $g$, at least for 
$g > g_{min}$ (as shown in Figs. S3(a) and (b) and explained in Text S1).

The reported results explain why the variability $\sigma(C)$ of the cross-correlation matrix has a non monotonic behaviour with $g$ (as shown in the middle panel in Fig.~\ref{fig:Q0andCV}(a)).
At low coupling $\sigma(C)$ is almost zero, since the single neuron dynamics
are essentially independent one from another, while the increase of the coupling leads to 
an abrupt rise of  $\sigma(C)$. \red{This growth is clearly associated with 
the inhibitory action which splits the neurons into correlated and anti-correlated groups.}
The variability of the cross-correlation matrix achieves a maximum value
for coupling slightly larger than $g_{min}$, where fluctuations in the effective
currents begin to play a relevant role in the network dynamics. At larger coupling
$\sigma(C)$ begins to decay towards a finite non zero value.
These results confirm that the most interesting region to examine   
is the one with coupling $g >  g_{min}$, as already suggested in ~\cite{ponzi2013optimal}.

The observed behaviour of $CV$, $n^*$ and $\sigma(C)$ suggests that we should expect 
a maximum in $Q_0$  at some intermediate coupling $g > g_{min}$, 
as indeed we have found for both
studied cases, as shown in Fig. \ref{fig:Q0andCV} (c) and (d). 
The initial increase in $Q_0$ is due to the increase in $CV$ and $n^*$, 
while the final decrease, following the occurrence of the maximum, 
is essentially driven by the decrease of $\sigma(C)$.
For larger $\Delta V$ the neurons tend to fire regularly in a wider range 
of coupling at small $g$ (see Fig. \ref{fig:Q0andCV} (d)), indicating that due to their higher firing
rates a larger synaptic inhibition is required to influence their dynamics.
On the other hand, their bursting activity observable
at large $g$ is more irregular 
(see the upper panel in Fig.~ \ref{fig:Q0andCV} (a); dashed line and empty symbols). 

\red{
To assess whether parameters that maximize $Q_0$ also allow discrimination between different inputs, 
we alternated the network back and forth between two distinct input patterns,
each presented for a period $T_{sw}$.}
\red{During this stimulation protocol, we evaluated the state transition matrix (STM) 
and the associated quantity $\Delta M_d$.
The STM measures the similarity among the firing rates of the neurons in the network 
taken at two different times. The metric $\Delta M_d$, based on the STM, 
has been introduced in~\cite{ponzi2013optimal} to quantify the 
ability of the network to distinguish between two inputs.
In particular, $\Delta M_d$ is the difference between the average values 
of the STM elements associated with the presentation of each of the two stimuli 
(a detailed description of the procedure is reported in the sub-section \textit{Discriminative and computation capability} and in Methods).
}

\red{
To verify whether the ability of the network to distinguish different stimuli is directly related to the presence of dynamically correlated and anti-correlated cell assemblies, we generated a new metric, $Q_d$. This metric is defined in the same way as $Q_0$, except that in  in Eq. (\ref{eq:Q0Indicator}) the standard deviation of the correlation matrix is replaced by $\Delta M_d$.
As it can be appreciated from Figs. \ref{fig:Q0andCV}(c) and \ref{fig:Q0andCV}(d)
the metrics $Q_d$ and $Q_0$ behave similarly, indicating that
indeed $Q_0$ becomes maximal in the parameter range in which 
the network is most effectively distinguishing different stimuli. 
We speculate that the emergence of correlated and anti-correlated assemblies contributes 
to this discriminative ability.}

\red{
We note that we observed maximal values of $Q_0$ for 
realistic lateral inhibition strengths, as measured from the 
post-synaptic amplitudes $A_{PSP}$. Specifically, 
$Q_0$ reaches the maximum at $g=4$ ($g=8$) for $\Delta V = 1$ mV ($\Delta V = 5$ mV)
corresponding to $A_{PSP} = 0.368$ mV ($A_{PSP} = 0.736$ mV), comparable to
previously reported experimental results~\cite{tunstall2002inhibitory,plenz2003,tepper2004gabaergic} (for more details see Methods).}

\subsection*{The role of the post-synaptic time scale}

\begin{figure}
\centering
\includegraphics[width=0.32\textwidth]{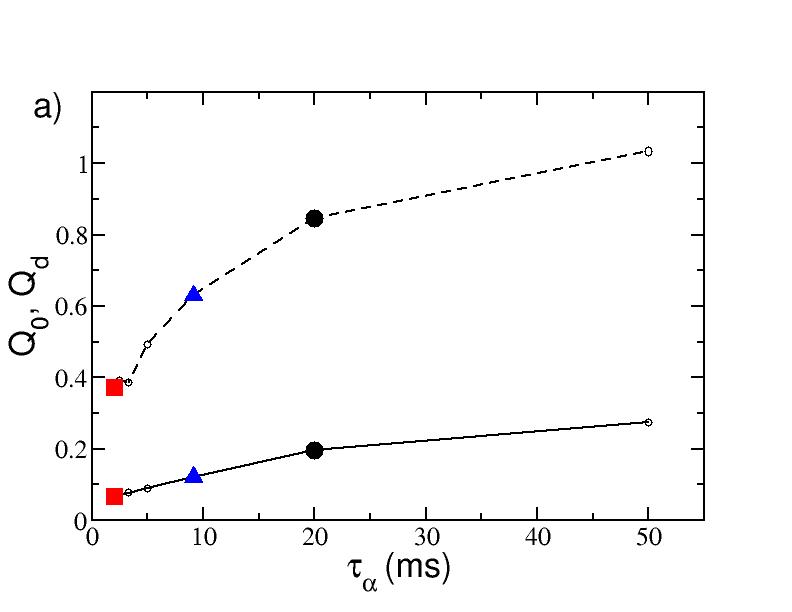}
\includegraphics[width=0.32\textwidth]{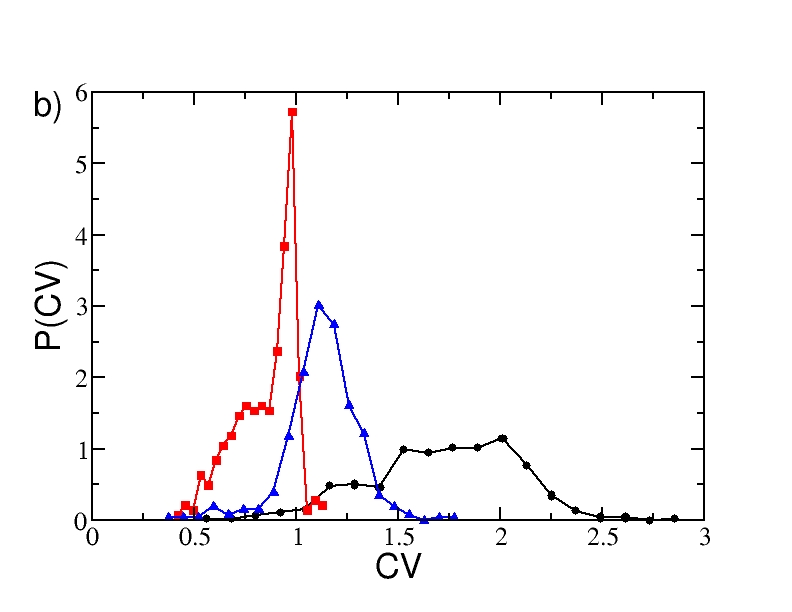}
\includegraphics[width=0.32\textwidth]{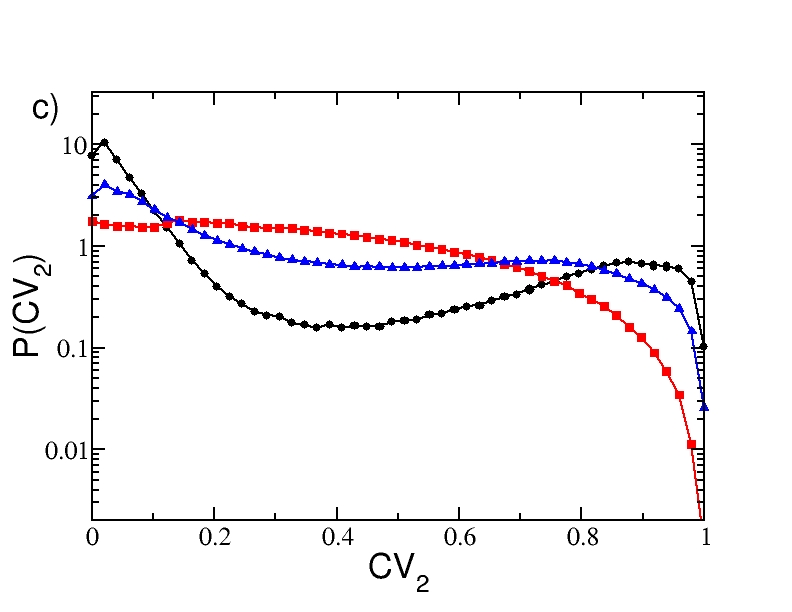}
\caption{\textbf{Metrics of structured activity vs post-synaptic time duration.} a) Metrics $Q_0$ (in solid line) and $Q_d$ (dashed) as a function of \red{the pulse time scale for the 
parameter values $\{\Delta V,g\} = \{5 \enskip {\rm mV},8\}$ corresponding to the maximum 
$Q_0$ value in Fig. \ref{fig:Q0andCV}(d)}. Probability distribution functions $P(CV)$ ($P(CV_2$)) for the coefficient of variation $CV$ (local coefficient of variation $CV_2$) are shown in b) (in c))
for three representative $\tau_{\alpha} = \{2, 9, 20\}$ ms, displayed
by employing the same symbols and colors as indicated in a). \red{For these three cases the average firing rate in the network is $\langle \nu \rangle = \{8.81, 7.65, 7.35\}$ Hz 
ordered for increasing $\tau_{\alpha}$-values}. 
Other parameters as in Fig. \ref{fig:PonziBenchmark}} 
\label{fig:CV1_CV2}
\end{figure}

\red{In brain slice experiments IPSCs/IPSPs between MSNs last 5-20 ms and are
mainly mediated by the GABA$_a$-receptor~\cite{tunstall2002inhibitory,koos2004comparison}. 
In this sub-section, we will examine the effect of the the post-synaptic time constant $\tau_\alpha$.
As $\tau_\alpha$ is increased from 2 to 50 ms, the values of of both metrics $Q_0$ and $Q_d$ 
progressively increase (Fig.~\ref{fig:CV1_CV2}(a)), with the largest variation having already occurred by $\tau_\alpha = 20$ ms. To gain more insights on the role of the PSP in shaping the structured dynamical regime, we show for the same network the distribution of the single cell $CV$, for  $\tau_{\alpha} = \{2, 9, 20\}$ ms (Fig. \ref{fig:CV1_CV2}(b)).}
Narrow pulses ($\tau_\alpha \simeq 2$ ms) are associated with a distribution of $CV$
values ranging from 0.5 to 1, with a predominant peak at one. By increasing $\tau_\alpha$ one observes that the $CV$ distributions shift to larger and larger $CV$ values.
Therefore, one can conclude that at small $\tau_\alpha$ the activity is mainly Poissonian,
\red{while increasing the duration of the PSPs leads to bursting behaviours,
as in experimental measurements of the MSN activity~\cite{miller2008dysregulated}. In particular in~\cite{miller2008dysregulated}, the
authors showed that bursting activity of MSNs with a distribution $P(CV)$ centered around 
$CV \simeq 2$ is typical of awake wild-type mice.}
To confirm this analysis we have estimated also the distribution of the $CV_2$: A $CV_2$ distribution with a peak around zero denotes a very regular firing, while a peak around one indicates the presence 
of long silent periods \red{followed by rapid firing events (i.e. a bursting activity).}
Finally a flat distribution denotes \red{Poissonian distributed spiking.}
It is clear from Fig.~\ref{fig:CV1_CV2}(c) that by increasing $\tau_\alpha$ from 2 to 20 ms 
this leads the system from an almost Poissonian behaviour to bursting dynamics, where 
almost regular firing inside the burst (intra-burst) is followed by a long quiescent 
period (inter-burst) before starting again.

\begin{figure*}
\centering
\includegraphics[width=0.47\textwidth]{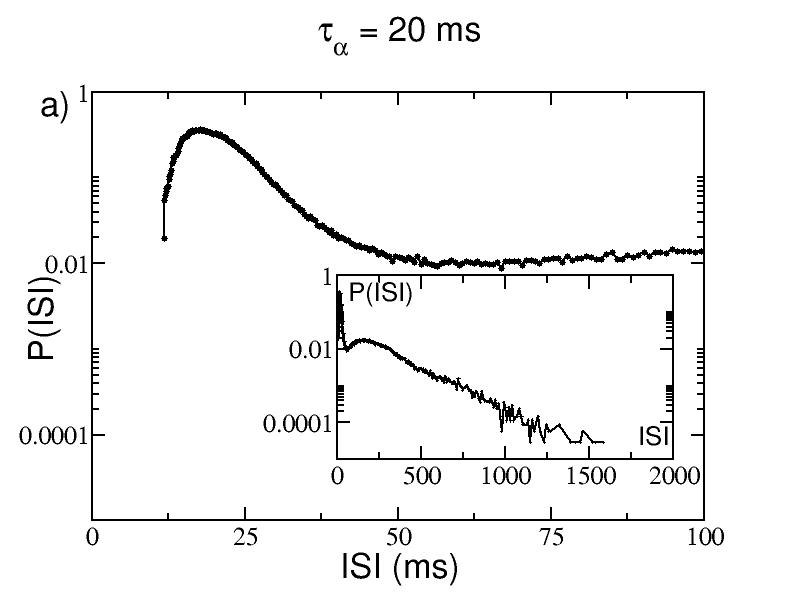}
\includegraphics[width=0.47\textwidth]{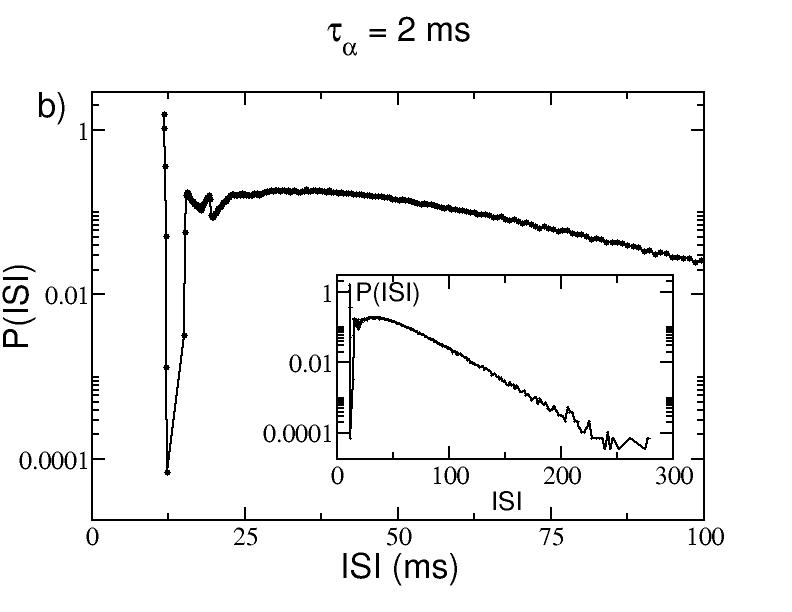}
\includegraphics[width=0.47\textwidth]{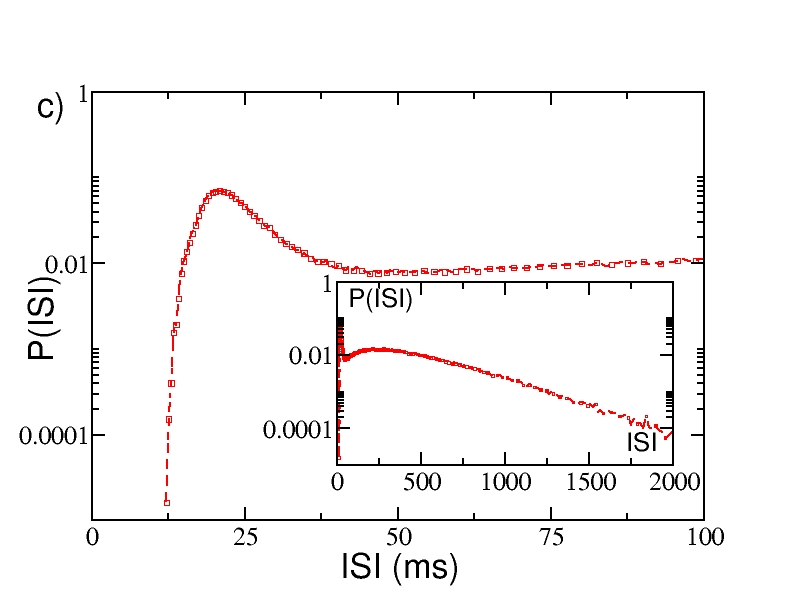}
\includegraphics[width=0.47\textwidth]{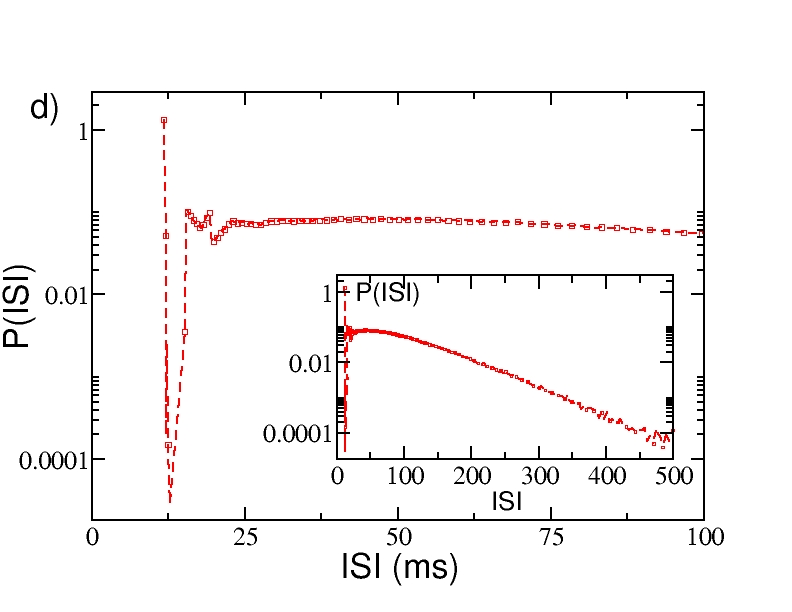}
\includegraphics[width=0.47\textwidth]{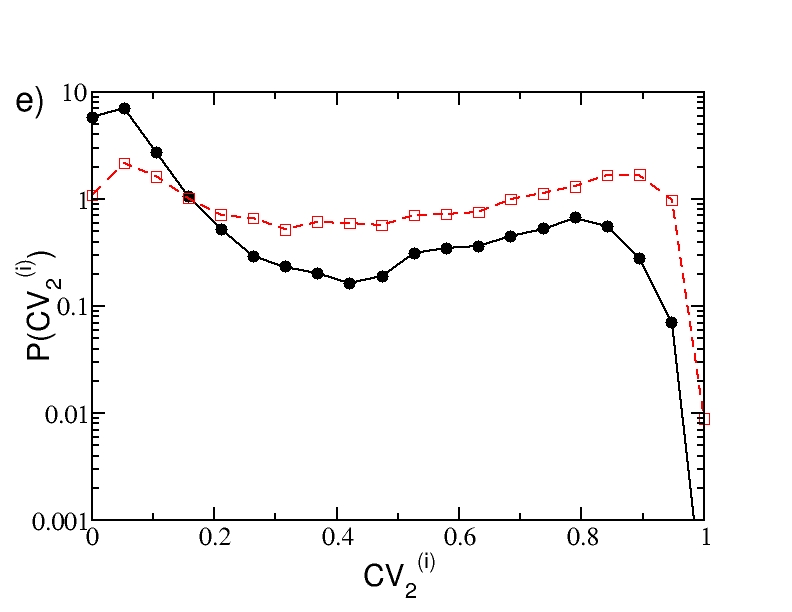}
\includegraphics[width=0.47\textwidth]{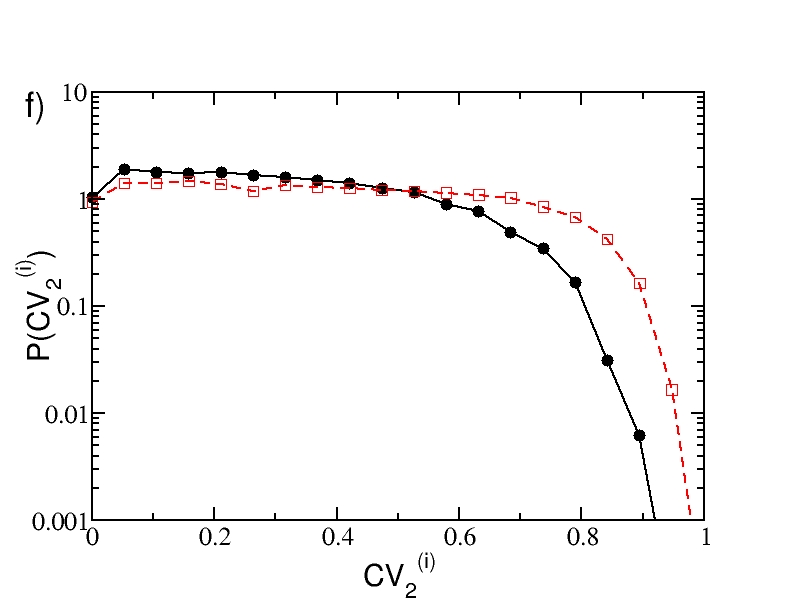}
\caption{\textbf{Single neuron statistics.} \red{First row : 
distributions $P(ISI)$ for one representative cell in the network are shown in black.
Second row: the corresponding Poissonian reconstruction of the $P(ISI)$ are reported in red.
In all plots the main figure displays the 
distributions at short ISIs, while the inset is a zoom out of the whole distribution.
Third row: single neuron distribution of the $CV^{(i)}_2$ for the 
considered neuron (black solid lines with circles) and its Poissonian distribution 
(red dashed line with  squares). The left (right) column 
corresponds to $\tau_\alpha = 20$ (2 ms).} 
The network parameters are $\Delta V = 5$ mV and $g = 8$, 
and the others as in Fig.~\ref{fig:PonziBenchmark}, both the examined neurons
have $I_s = -45.64$ mV. For the Poissonian reconstruction the frequencies of the incoming 
uncorrelated spike trains are set to \red{$\langle \nu \rangle_N \approx 7.4$ Hz ($\langle \nu \rangle_N \approx 8.3$ Hz)} for $\tau_{\alpha} = 20$ms ($\tau_{\alpha} = 2$ms), as  measured from the corresponding network dynamics. The distributions are obtained by considering a sequence of $10^9$ spikes in the original network, and $10^7$ events for the Poissonian reconstruction. 
}
\label{fig:ISI_distribut}
\end{figure*}

The distinct natures of the distributions of $CV$ for short and long-tailed pulses
raises the question of what mechanism underlies such differences. 
To answer this question we analyzed the distribution of the ISI of a single cell in the network 
for two cases: in a cell assembly bursting regime (corresponding to $\tau_\alpha = 20$ ms) and 
for Poissonian unstructured behavior (corresponding to $\tau_\alpha = 2$ ms).
\red{We expect that even the single neurons should have completely different dynamics
in these two regimes, since the distributions $P(CV)$ at $\tau_\alpha = 2$ ms 
and 20 ms are essentially not overlapping, as shown in Fig. \ref{fig:CV1_CV2}(b).
In order to focus the analysis on the effects due to the synaptic inhibition,
we have chosen, in both cases, neurons receiving exactly the same external excitatory drive
$I_s$. Therefore, in absence of any synapses, these two neurons will fire with the same period
ISI$_0 = \tau_m \log [(I_s-V_{r})/(I_s-V_{th})] = 12$ ms, corresponding to a firing rate of 
8.33 Hz not far from the average firing rate of the networks (namely, $\langle \nu \rangle_N \simeq 7-8$ Hz). Thus these neurons can be considered as displaying a typical activity in both regimes.
As expected, the dynamics of the two neurons is quite different,
as evident from the $P(ISI)$ presented in Fig. \ref{fig:ISI_distribut}(a) and (b).}
In both cases one observes a long tailed exponential decay
of $P(ISI)$ corresponding to a Poissonian like behaviour. 
However the decay rate $\nu_D$ is slower for $\tau_\alpha = 20$ ms with respect to $\tau_\alpha = 2$ ms,
namely $\nu_D \simeq 2.74$ Hz versus $\nu_D \simeq 20.67$ Hz.
Interestingly, the main macroscopic differences between the two distributions
arises at short time intervals. For $\tau_{\alpha} = 2$ ms, (see Fig. \ref{fig:ISI_distribut}(b))
an isolated and extremely narrow peak appears at  ISI$_0$.
\red{This first peak corresponds to the supra-threshold tonic-firing of the isolated neuron,
as reported above. After this first peak, a gap is clearly visible in the $P(ISI)$ followed by
an exponential tail. The origin of the gap resides in the fact that ISI$_0 >> \tau_\alpha$,
because if the neuron is firing tonically with its period ISI$_0$ and receives a single PSP, 
the membrane potential has time to decay almost to the reset value $V_r$ 
before the next spike emission. Thus a single PSP will delay the next firing event
by a fixed amount corresponding to the gap in Fig.~\ref{fig:ISI_distribut}(b). 
Indeed one can estimate analytically this delay due to the arrival
of a single $\alpha$-pulse, in the present case this gives ISI$_1$ = 15.45 ms, in very
good agreement with the results in Fig. \ref{fig:ISI_distribut}(b). 
No further gaps are discernible in the distribution, because it is highly improbable that 
the neuron will receive two (or more) PSPs exactly at the same moment at reset, 
as required to observe further gaps.
The reception of more PSPs during the ramp up phase will give rise to
the exponential tail in the $P(ISI)$}. In this case the contribution to the $CV$ comes essentially from this exponential tail, 
while the isolated peak at ISI$_0$ has a negligible contribution.

 On the other hand, if $\tau_\alpha > \text{ISI}_0$, as in the
case reported in Fig. \ref{fig:ISI_distribut}(a), $P(ISI)$ does not show
anymore a gap, but instead a continuous distribution of values. This because now the 
inhibitory effects of the received PSPs sum up leading to a continuous range of
delayed firing times of the neuron. \red{The presence of this peak of finite width 
at short $ISI$ in the $P(ISI)$ plus the exponentially decaying tail are at the origin 
of the observed $CV > 1$. In Fig. \ref{fig:ISI_distribut} (e) and \ref{fig:ISI_distribut} (f) the  distributions of the coefficient $CV^{(i)}_2$ are also displayed for the considered neurons as black
lines with symbols. These distributions clearly confirm that the dynamics are bursting for the longer synaptic time scale and essentially Poissonian for the shorter one.}

We would like to understand whether it is possible to reproduce similar distributions of 
the ISIs by considering an isolated cell receiving Poissonian distributed inhibitory inputs. 
In order to verify this, we simulate a single cell receiving $K$ uncorrelated spike trains 
at a rate $\langle \nu \rangle_N$, or equivalently, a single Poissonian spike train 
with rate $K \langle \nu \rangle_N$. Here, $\langle \nu \rangle_N$ is the average 
firing rate of a single neuron in the original network. The corresponding $P(ISI)$ 
are plotted in Fig.~\ref{fig:ISI_distribut} (c) and \ref{fig:ISI_distribut} (d), for $\tau_\alpha = 20$ ms and 2 ms, respectively. There is a remarkable similarity between the reconstructed ISI distributions and the real ones (shown in Fig. \ref{fig:ISI_distribut}(a) and (b)) , in particular at short ISIs.
\red{Also the distributions of the $CV^{(i)}_2$ for the reconstructed dynamics are
similar to the original ones, as shown in Fig. \ref{fig:ISI_distribut} (e) and \ref{fig:ISI_distribut} (f). Altogether,} these results demonstrate that the bursting activity of inhibitory coupled cells is not a consequence of complex correlations among the incoming spike trains, but rather a characteristic
related to intrinsic properties of the single neuron: \red{namely, its tonic firing period}, the synaptic strength, and the post-synaptic time decay. \red{The fundamental role played by long synaptic time 
in inducing bursting activity has been reported also in a study of a single LIF neuron 
below threshold subject to Poissonian trains of exponentially decaying PSPs~\cite{moreno2004}.
}

\red{
Obviously this analysis cannot explain collective effects, like the 
non trivial dependence of the number of active cells $n^*$ on the synaptic
strength, discussed in the previous sub-section, or the emergence of correlations and anti-correlations among neural assemblies (measured by $\sigma(C)$) due to the covarying of the firing rates in the network, as seen in the striatum slices and shown in Fig. \ref{fig:PonziBenchmark} (c) for our model.
To better investigate the influence of $\tau_\alpha$ on the collective properties of the
network we report in Fig. S5(a) and (b) the averaged CV, $\sigma(C)$, $n^*$ and $\Delta M_d$ for 
$\tau_\alpha \in [2,50]$ ms. As already noticed, the network performs better
in mimicking the MSN dynamics and in discriminating between different inputs
at larger $\tau_\alpha$ (e.g. at 20 ms). However, what is the minimal 
value of $\tau_\alpha$ for which the network still reveals cell assembly dynamics and 
discriminative capabilities ?
From the data shown in Fig. S5(a) one can observe that $\sigma(C)$ 
and $\Delta M_d$ attain their maximal values in the range 10 ms $\le \tau_\alpha \le$ 20 ms. 
This indicates that clear cell assembly dynamics with associated good discriminative skills
can be observed in this range. However, the bursting activity is not 
particularly pronounced at $\tau_\alpha = 10 $ ms, 
where $\langle CV \rangle_N \simeq 1$. Therefore only the 
choice $\tau_\alpha = 20 $ ms fulfills all the requirements.
}

\red{Interestingly, genetic mouse models of Huntington's disease (HD)
revealed that spontaneuous IPSCs in MSNs has a reduced decay time and half-amplitude
duration compared to wild-types~\cite{cummings2010}.
Our analysis clearly indicate that a reduction of $\tau_\alpha$
results in more stochastic single-neuron dynamics,
as indicated by $\langle CV \rangle_N \simeq 1$, as well as in
a less pronounced structured assembly dynamics (Fig. S5 (a)). 
This resembles what observed for the striatum dynamics of freely behaving mice with 
symptomatic HD~\cite{miller2008dysregulated}. In particular,
the authors have shown in~\cite{miller2008dysregulated} that at the single unit level
HD mice reveals a $CV \simeq 1$ in contrast to $CV \simeq 2$ for wild-type mice,
furthermore the correlated firing was definitely reduced in HD mice.
}

\red{\subsection*{Structural origin of the cell assemblies}}

\begin{figure*}
\centering
\includegraphics[width=0.4\textwidth]{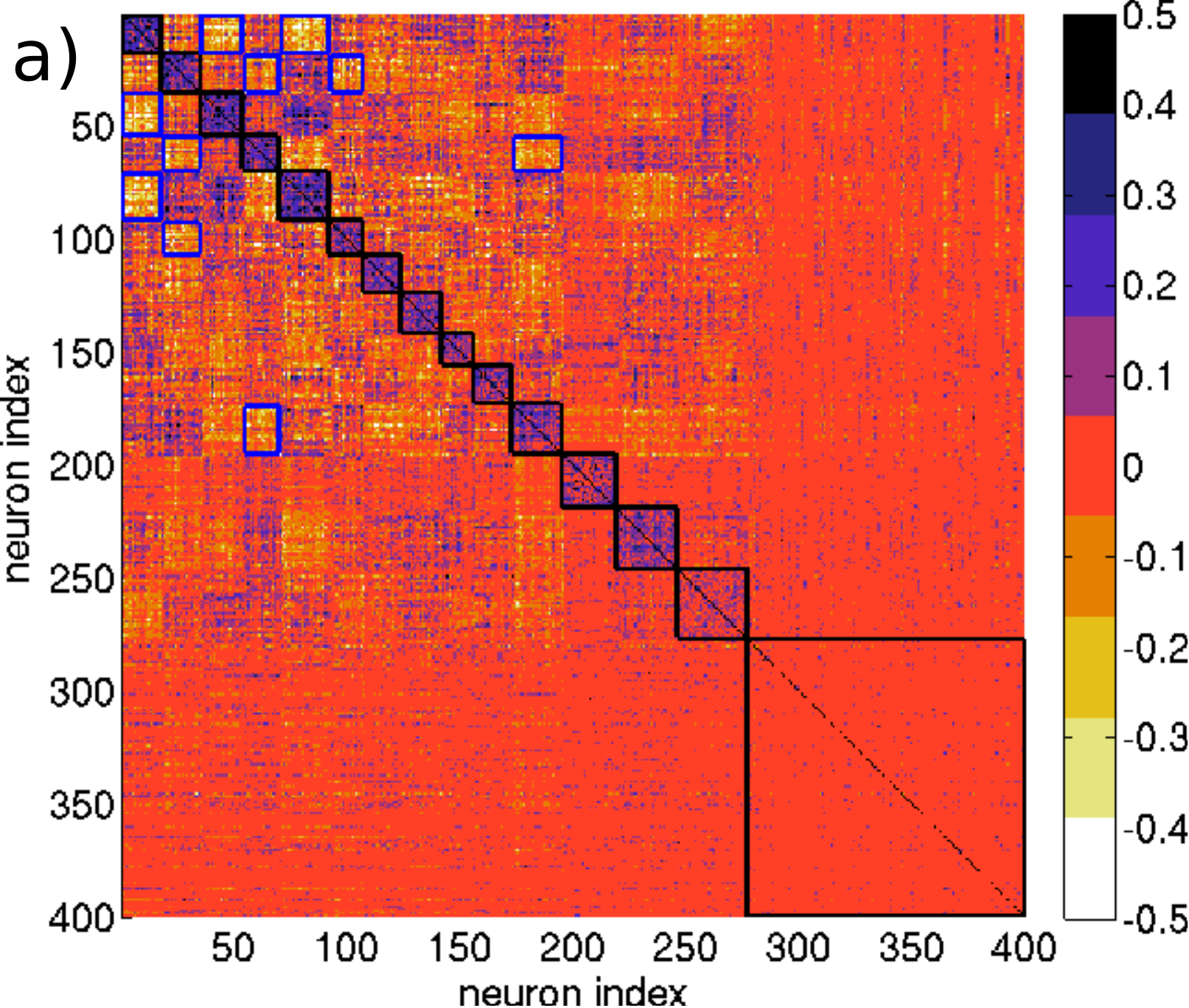}
\includegraphics[width=0.36\textwidth]{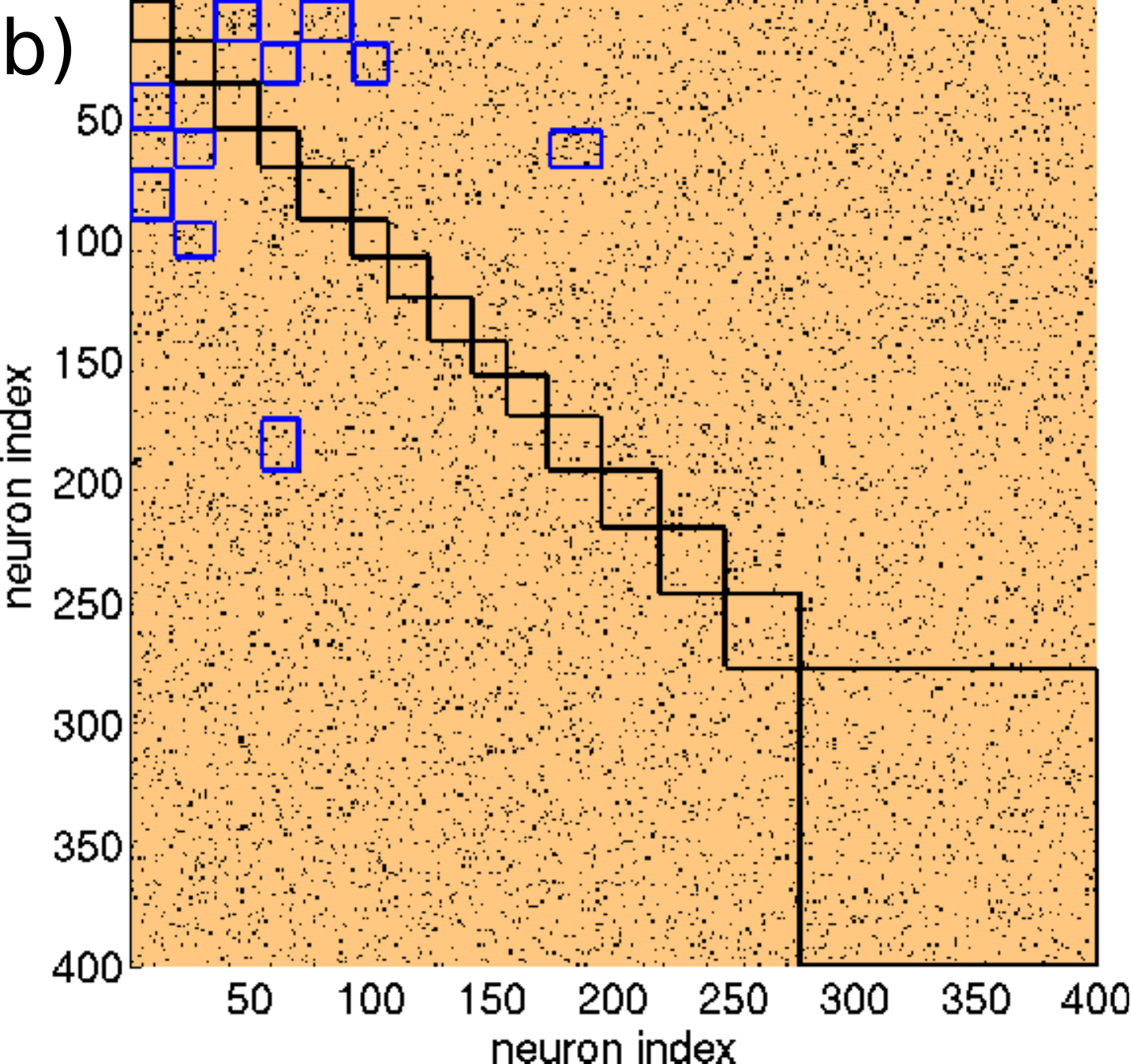}\\
\includegraphics[width=0.4\textwidth]{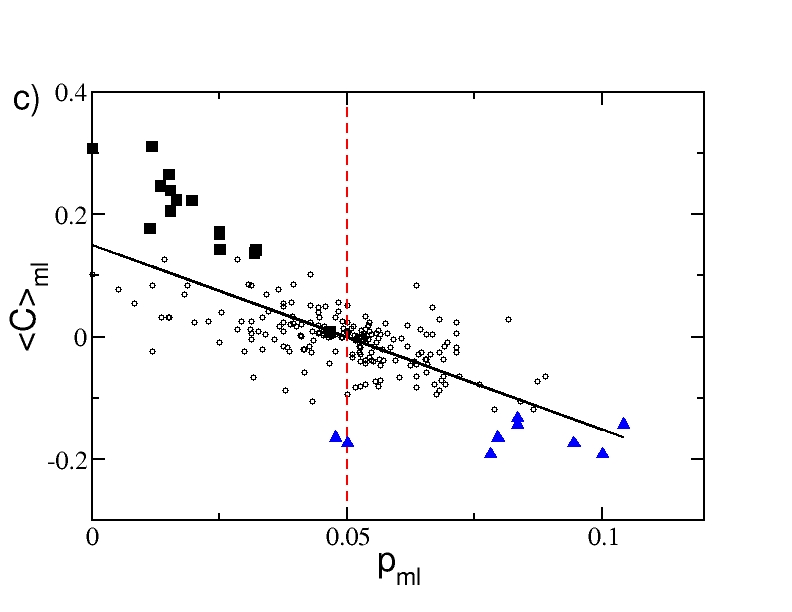}
\caption{\red{\textbf{Cell assemblies and connectivity.} a) Cross-correlation matrix $C(\nu_i,\nu_j)$ of the firing rates organized according to the clusters generated via the \textit{k-means} algorithm with $k=15$, the clusters are ordered as in Fig.~\ref{fig:PonziBenchmark}(c) from the highest to the lowest correlated one. b) Connectivity matrix $\mathcal{C}_{ij}$ with the indices ordered as in panel a). Here, a black (copper) dot denotes a 1 (0) in $\mathcal{C}_{ij}$, i.e. the presence of a synaptic connection from
$j$ to $i$. c) Average cross-correlation $\langle C \rangle_{ml}$ among the elements of the matrix block $(m,l)$ , versus the probability
$p_{ml}$ to have synaptic connections from neurons in the cluster $l$ to neurons in the cluster $m$.
Black squares indicate the blocks along the diagonal delimited by black borders in panel a) and b) , i.e. they correspond to the pairs $\{ \langle C \rangle_{mm}, p_{mm} \}$; blue triangles denote the ten blocks with the lowest $\langle C \rangle_{ml}$ values, which are also delimited by blue edges in a) and b). The vertical red dashed line indicates the average probability to have a connection $p = 5 \%$. The black solid line is the linear regression 
to the data ($ \langle C \rangle_{ml} \approx 0.15 - 3.02  p_{ml}$, correlation coefficient  $R = -0.72$). Other parameters as in Fig. \ref{fig:PonziBenchmark}.}}
\label{fig:structural}
\end{figure*}

\red{
A question that we have not addressed so far is: how do cell assemblies arise ?
Since the network is purely inhibitory it is reasonable to guess that correlation (anti-correlation) among groups of neurons will be related to the absence (presence) of synaptic connections between the considered groups. In order to analyze the link between the correlation and the network connectivity
we compare the clustered cross-correlation matrix of the firing rates 
$C(\nu_i,\nu_j)$ (shown in Fig \ref{fig:structural} (a))
with the associated connectivity matrix $\mathcal{C}_{ij}$ (reported in Fig-~\ref{fig:structural} (b)).
The cross-correlation matrix is organized in $k=15$ clusters via the {\it $k$-means} algorithm,
therefore we obtain a matrix organized in a $k \times k$ block structure, where each block $(m,l)$ contains all the cross-correlation values of the elements in cluster $m$ with the elements in cluster $l$. The connectivity matrix is arranged in exactly the same way, however it should be
noticed that while $C(\nu_i,\nu_j)$ is symmetric, the matrix $\mathcal{C}_{ij}$  is not symmetric due to the unidirectional nature of the synaptic connections. From a visual comparison of the two figures it is clear that the most correlated blocks are along the diagonal and that the number of connections
present in these diagonal blocks is definitely low, with respect to the expected value from the whole matrix. An
exception is represented by the largest diagonal block which reveals, however, an almost zero level of correlation
among its members. We have highlighted in blue some blocks with high level of anti-correlations among the elements, the same blocks in the connectivity matrix reveal a high
number of links. A similar analysis, leading to the same conclusions was previously reported 
in ~\cite{ponzi2010sequentially}.}

\red{
However, we would like to make more quantitative this comparison. Therefore we have estimated
for each block the average cross-correlation, denoted as $\langle C \rangle_{ml}$, and
the average probability $p_{ml}$ of unidirectional connections from the cluster $l$ to the cluster $m$.
These quantities are shown in Fig.~\ref{fig:structural} (c) for all
the possible blocks, it is evident that the correlation $\langle C \rangle_{ml}$
decreases with the probability $p_{ml}$, a linear fit to the data is reported
in the figure as a solid black line. However, there are blocks that are outliers with
respect to this fit, in particular the black squares refer to the diagonal blocks
and these are all associated to high correlation values $\langle C \rangle_{mm}$
and low probabilities $p_{mm}$, definitely smaller than the average probability $p=0.05$,
shown as a dashed vertical red line in Fig.~\ref{fig:structural} (c).
An exception is represented by a single black square located exactly on the linear fit
in proximity of $p=0.05$, this is the large block with almost zero level of correlation
among its elements previously identified. Furthermore, the blocks with higher anticorrelation,
denoted as blue triangles in the figure, have probabilities $p_{ml}$ definitely larger
than 5 \%. Also in this case there are 2 exceptions, 2 triangles lie exactly on the vertical dashed line corresponding to 5 \%. This is due to the fact that the $p_{ml}$ are not symmetric, and it
is sufficient to have a large probability to have connections in only one of
the two possible directions between blocks $m$ and $l$ to observe anti-correlated activities between the two assemblies. To summarize we have clearly shown that the origin of the assemblies
dynamically identified from the correlations of the firing rates is directly related to 
structural properties of the networks, as visualized by the connectivity matrix.}\\

\subsection*{Discriminative and computational capability} 

In this sub-section we examine the ability of the network to perform different tasks: namely, to respond in a reproducible manner to equal stimuli and to discriminate between similar inputs 
via distinct dynamical evolution. For this analysis we have always compared the responses of the network obtained for a set of parameters corresponding to the maximum $Q_0$ value shown in Fig.~\ref{fig:Q0andCV}(d),
where $\tau_{\alpha} = 20$ ms, and for the same parameters but with a shorter PSP decay time,
namely $\tau_{\alpha} = 2$ ms.

\begin{figure*}
\centering
\includegraphics[width=0.47\textwidth]{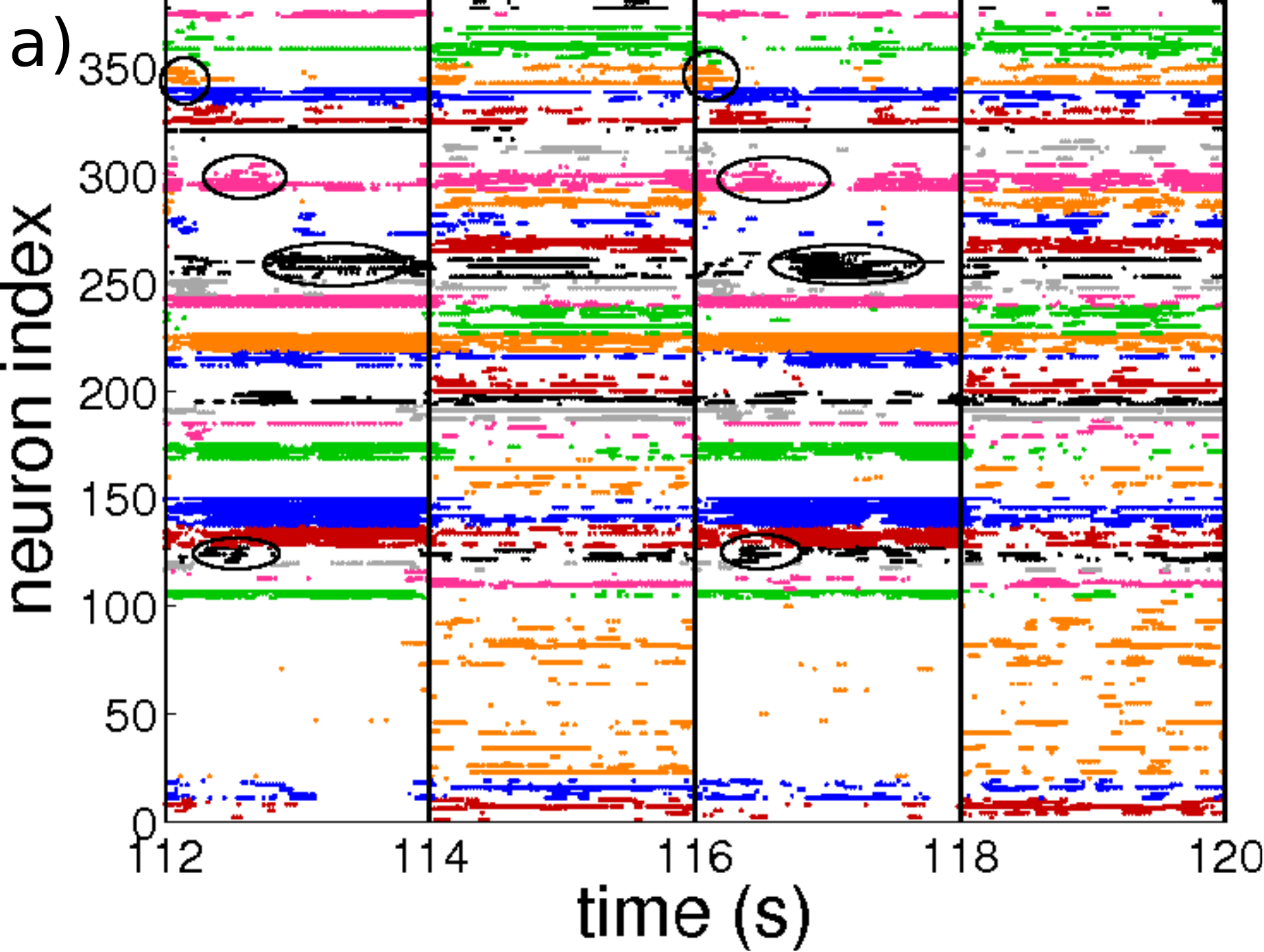}
\includegraphics[width=0.47\textwidth]{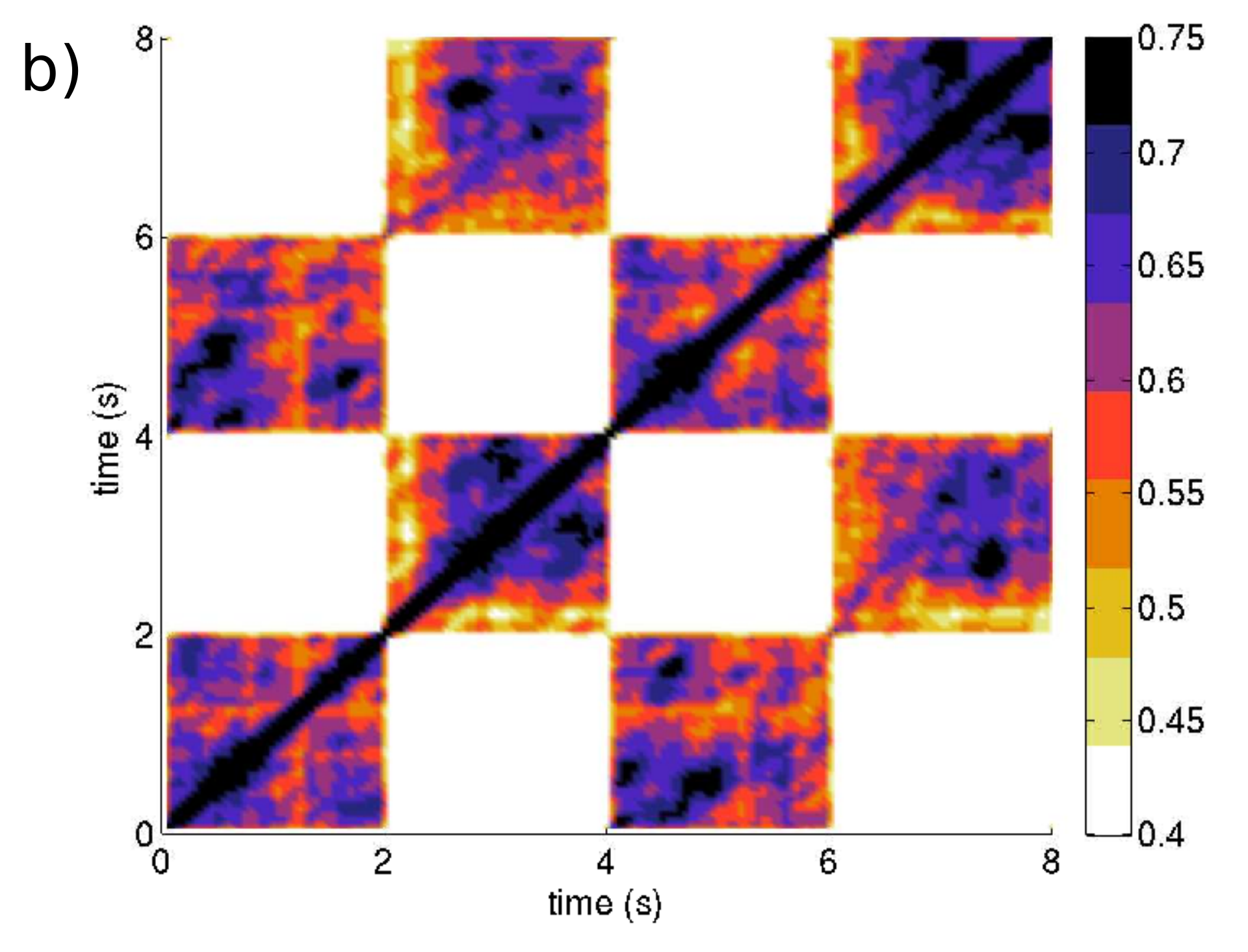}
\caption{\textbf{Sequential switching.} a) Raster plot associated to the two input protocols $I^{(1)}$ and $I^{(2)}$. The circles denote the clusters of active neurons appearing repetitively after the presentation of the stimulus $I^{(1)}$. Vertical lines denote the switching times between stimuli. b) Averaged State Transition Matrix $\overline{D}$ , \red{obtained by considering a $4 T_{sw} \times 4 T_{sw}$ sub-matrix averaged over $r=5$ subsequent time windows of duration $4 T_{sw}$} (see the section \textit{Methods} for details). The inputs $I^{(1)}$ and 
$I^{(2)}$ are different realization of the 
same random process, they are obtained by selecting $N$ current values $I_i$ from the flat interval $[V_{th}, \, V_{th}+\Delta V]$. The input stimuli are switched every $T_{sw} = 2$ s. Number of clusters $k = 35$ in a). Other parameters as in Fig. \ref{fig:PonziBenchmark}.}
\label{fig:SequantialSwitching}
\end{figure*}

To check for the capability of the network to respond to cortical inputs with a reproducible sequences of states of the network, we perform a simple experiment where two different inputs $I^{(1)}$ and $I^{(2)}$ are presented sequentially to the system. Each input persists for a time duration $T_{sw}$  and then the stimulus is switched to the other one and this process is repeated for the whole simulation time. The raster plot measured during
such an experiment is shown in Fig.~\ref{fig:SequantialSwitching} (a) for $\tau_{\alpha} = 20$ ms.
Whenever one of the stimuli is presented, a specific sequence of pattern activations can be observed. Furthermore, the sequence of emerging activity patterns is reproducible 
when the same stimulus is again presented to the system, as can be appreciated
by observing the patterns encircled with black lines in Fig.~\ref{fig:SequantialSwitching} (a).
Recall that the clustering algorithm here employed to identify the different groups 
is applied only during the presentation of the first stimulus,
therefore the sequential dynamics is most evident for that particular stimuli.
 
Furthermore, we can quantitatively calculate how similar is the firing activity in the network 
at different times by estimating the STM. The similarity is 
quantified by computing the normalized scalar product of the instantaneous firing rates of 
the $N$ neurons measured at time $t_i$ and $t_j$. We observe that
the similarity of the activity at a given time $t_0$ and at a successive time 
$t_0 + 2 m T_{sw}$ is high \red{(with values between 0.5 and 0.75), thus suggesting that the response 
to the same stimulus is similar, 
while it is essentially uncorrelated with the response 
at times corresponding to the presentation of a different stimulus, i.e. at $t_0 + (2m-1) T_{sw}$ (since the  similarity is always smaller than 0.4)} (here, $m = 1,2,3...$). This results in a STM with a periodic structure of period $T_{sw}$ with alternating high correlated blocks followed by low correlated blocks (see Fig. S6(b)). An averaged version of the STM \red{calculated over a sequence of 5 presentations of $I^{(1)}$ and $I^{(2)}$ is shown in Fig.~\ref{fig:SequantialSwitching} (b) (for details of the calculation see Methods). These results show not only the capability of the network to distinguish between the stimuli, but also the reproducible nature of the system response. In particular, from Fig.~\ref{fig:SequantialSwitching} (b) it is evident how the patterns associated with the response to the stimulus $I^{(1)}$ or $I^{(2)}$ are clearly different and easily identifiable. We also repeated the numerical experiment for another 
different realization of the inputs, noticing essentially the same features previously reported 
(as shown in Fig. S6(a-c)).} \red{
Furthermore, to test for the presence of memory effects influencing
the network response, we performed a further test where the system
dynamics was completely reset after each stimulation and before the
presentation of the next stimulus. We do not observe any relevant change 
in the network response, so we can conclude that our results are robust.}

\red{
Next, we examined the influence of the PSP time scale on the observed results.
In particular, we considered the case $\tau_{\alpha} = 2$ ms, for this value
the network does not reveal a large variability in the response showing a 
reduced number of patterns of activity.
In particular, as  shown in Fig. S6(d) it responds in a quite uniform manner during 
the presentation of each stimulus. Furthermore, the corresponding STM
reported in Figs.~S6(e) shows highly 
correlated blocks alternating to low correlated ones, but these blocks do not 
reveal any internal structure typical of cell assembly encoding.}

\begin{figure}
\centering
\includegraphics[width=0.47\textwidth]{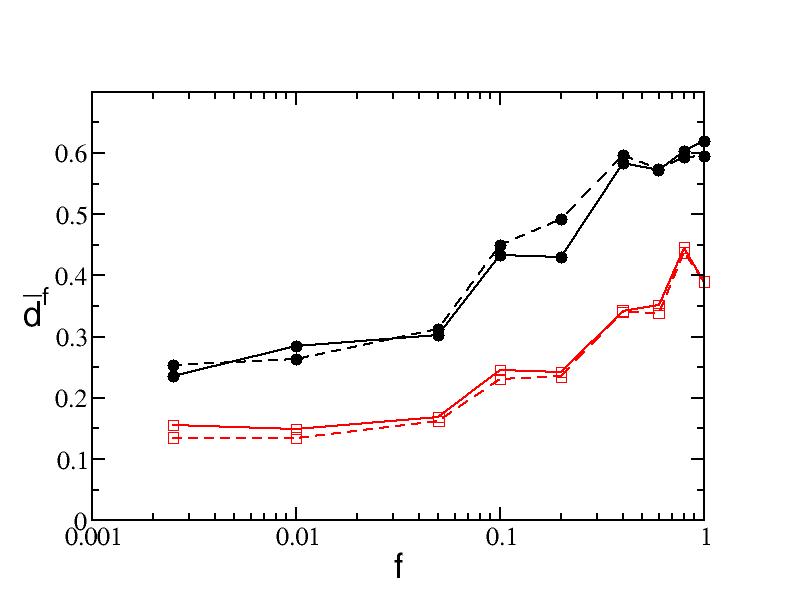}
\caption{{\bf Pattern separation.} \red{Average  dissimilarity as a function of the fraction $f$ of inputs differing from the control input,} for the values of $\tau_{\alpha} = 20$ms (black circles) and $\tau_{\alpha} = 2$ms (red squares) with two different observation windows $T_{E} = 2$s (solid line) and $T_{E} = 10$s (dashed line). Other parameters used: $\Delta T= 50$ms, $\Delta V= 5$ mV. Remaining parameters as in Fig. \ref{fig:PonziBenchmark}.}
\label{fig:Pattern_Sep}
\end{figure}

We proceeded to check the ability of the network to discriminate among similar
inputs and how this ability depends on the temporal scale of the synaptic response.
In particular, we tried to answer to the following question:
if we present two inputs that differ only for a fraction $f$ of the stimulation
currents, which is the minimal difference between the inputs that the network
can discriminate ?  In particular, we considered a control stimulation $I^{(c)} = {I_i} \in [V_{th}, \, V_{th}+\Delta V]$ 
and a perturbed stimulation $I^{(p)}$, where the stimulation currents
differ only over a fraction $f$ of currents $I_i$ (which are randomly chosen from the same 
distribution as the control stimuli). We measure the differences of the responses
to the control and to the perturbed stimulations by measuring, over an observation window $T_E$,
the dissimilarity metric $d^f(t)$, defined in Methods. The time averaged dissimilarity metric
$\bar d^f$ is reported as a function of $f$ in Fig.~\ref{fig:Pattern_Sep}
for two different values $\tau_{\alpha}$. It is clear that for any $f$-value 
the network with longer synaptic response always discriminates better between the
two different stimuli than the one with shorter PSP decay. We have also verified that the metric 
is robust to the modification of the observation times $T_E$, this is verified
because the dissimilarity $d^f(t)$ rapidly reaches a steady value (as shown in Fig. S7(a) and (b)).

In order to better characterize the computational capability of the network 
and the influence due to the different duration of the PSPs, 
\red{we measure the complexity of the output signals as recently suggested in~\cite{ostojic2014two}.
In particular, \red{we have examined the response of the network to a sequence of three 
stimuli, each being a constant vector of randomly chosen currents.}
The three different stimuli are consecutively presented to the network
for a time period $T_{sw}$, }
and the stimulation sequence is repeated for the whole experiment duration $T_{E}$. 
The output of the network can be represented by the instantaneous firing rates
of the $N$ neurons measured over a time window $\Delta T = 100$ ms,  this is a
high dimensional signal, where each dimension is represented by the activity
of a single neuron. \red{The complexity of the output signals can be estimated by measuring
how many dimensions are explored in the phase space, more stationary are
the firing rates less variables are required to reconstruct the whole output signal~\cite{ostojic2014two}.}

\begin{figure}
\centering
\includegraphics[width=0.32\textwidth]{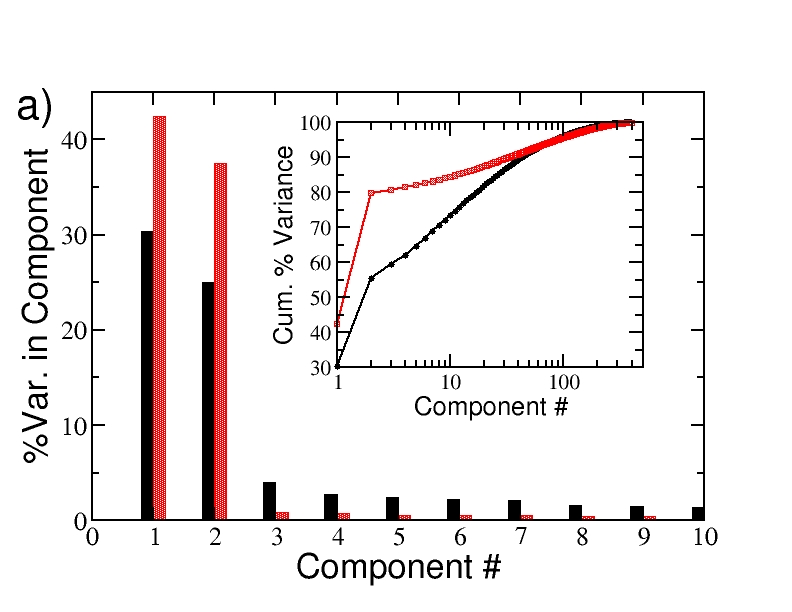}
\includegraphics[width=0.32\textwidth]{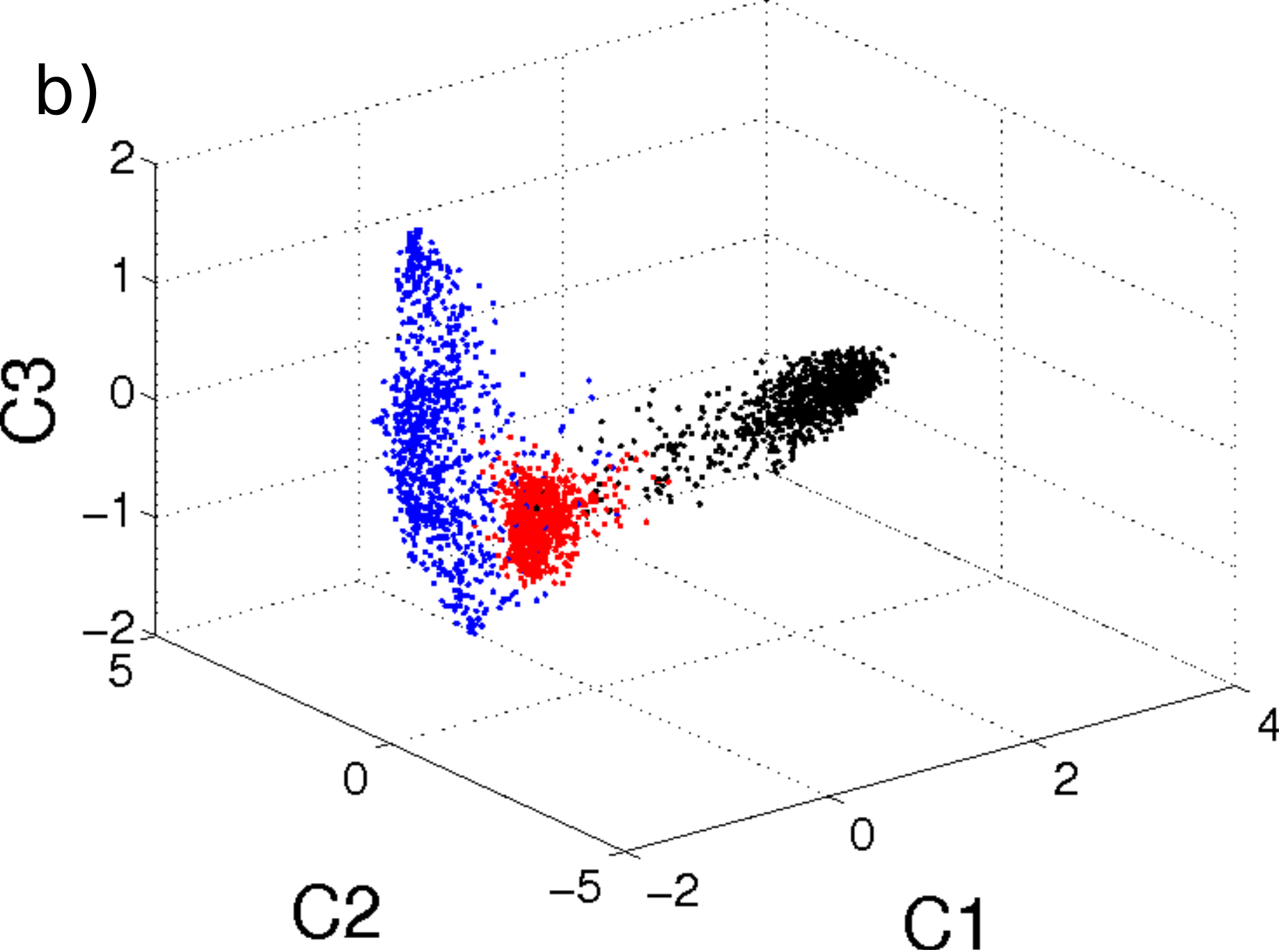}
\includegraphics[width=0.32\textwidth]{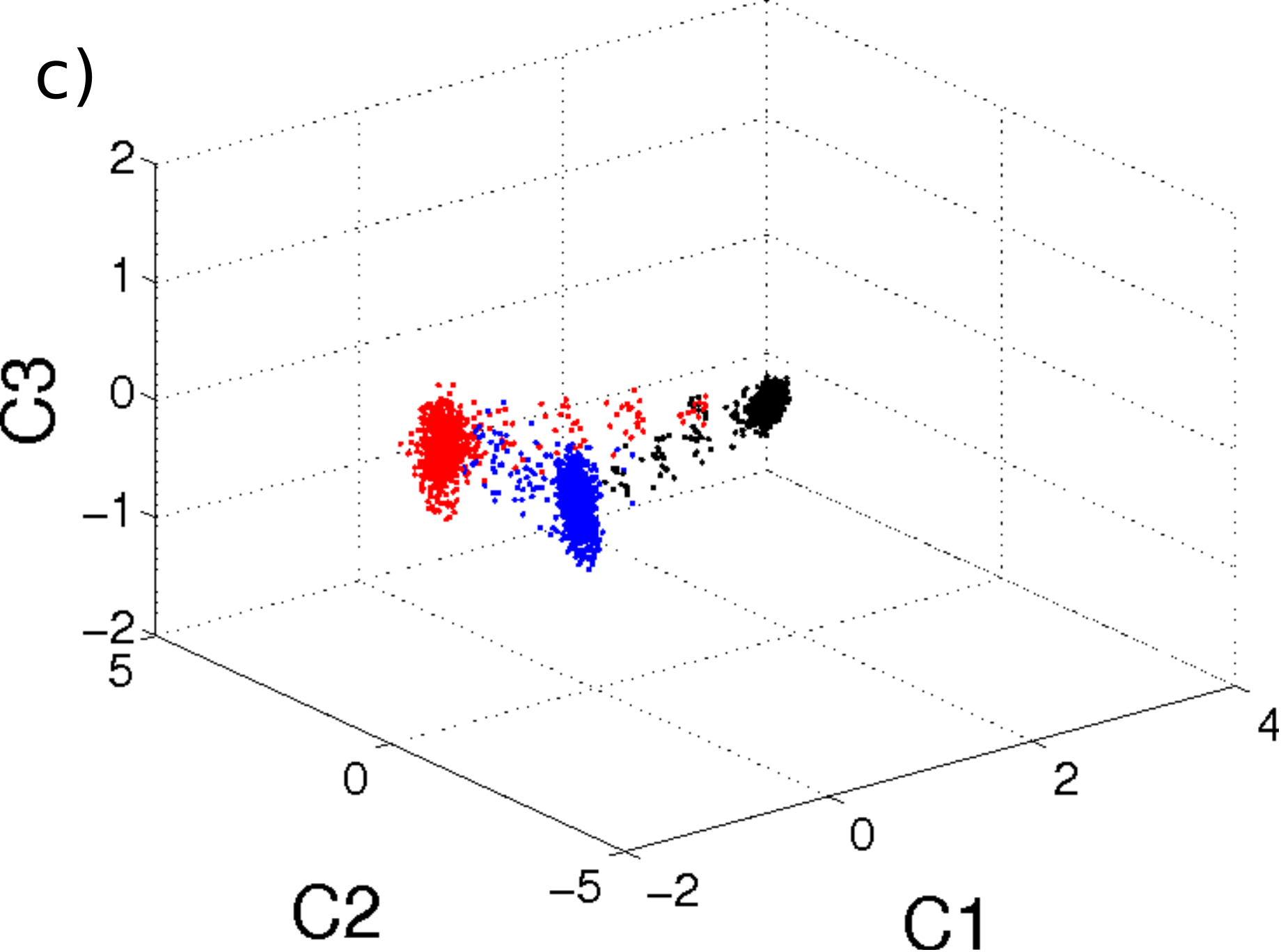}
\caption{\textbf{Computational capability of the network.} 
Characterization of the firing activity of the network, obtained as response to three consecutive 
inputs presented in succession.
a) Percentage of the variance of the neuronal firing activity 
reproduced by each of the first 10 principal components. The inset displays the corresponding cumulative percentage as a function of the considered component. Filled black and shaded red (bar or symbols) correspond to $\tau_{\alpha}$ = 20 ms and $\tau_{\alpha}$ = 2 ms, respectively. 
Projection of the neuronal response along the first three principal components
for b) $\tau_{\alpha}$ = 20 ms and c) $\tau_{\alpha}$ = 2 ms. \red{Each point in 
the graph correspond to a different time of observation.}
The three colors denote the response to the three different inputs,
which are constant stimulation currents randomly taken as $I^{(j)} \in [V_{th}, \, V_{th} + \Delta V]$ 
for $j = 1,2,3$, the experiment is then performed as explained in the text.}
\label{fig:PCA_200}
\end{figure}

A  principal component analysis (PCA) performed over $T_{E}/\Delta T$ 
observations of the $N$ firing rates reveals that for $\tau_{\alpha} = 2$ ms
the 80\% of the variance is recovered already with a projection over a two dimensional
sub-space (red bars in Fig.~\ref{fig:PCA_200} (a)). On the other hand, for $\tau_{\alpha} = 20$ ms
a higher number of principal components  is required to reconstruct the dynamical evolution (black
bars in Fig.~\ref{fig:PCA_200} (a)), thus suggesting higher computational capability of the system with longer PSPs~\cite{ostojic2014two}.  

These results are confirmed by analyzing the projections of the firing rates in 
the subspace spanned by the first three principal components $(C1,C2,C3)$ 
shown in Fig.~\ref{fig:PCA_200} (b) and (c) for $\tau_{\alpha} = 20$ ms and 
$\tau_{\alpha} = 2$ ms, respectively. The responses to the three different stimuli 
can be effectively discriminated by both networks, since they lie in different parts of 
the phase space. However, the response to the three stimuli correspond essentially to three fixed
points for $\tau_{\alpha} = 2$ ms, while trajectories evolving in a higher dimension are 
associated to each constant stimulus for $\tau_{\alpha} = 20$ ms. 
 
\red{These analyses confirm that the network parameters selected by employing the maximal $Q_0$ criterion also result in a reproducible response  
to different stimuli, as well as in an effective discrimination between different inputs.}

\red{
In a recent work Ponzi and Wickens~\cite{ponzi2013optimal} have noticed that
in their model the striatally relevant regimes correspond to marginally stable dynamical evolution.
In the Supporting Information Text S1 we devote the sub-section {\it Linear stability analysis} to the
investigation of this specific point, our conclusion is that for our model the
striatally relevant regimes are definitely chaotic, but located
in proximity of a transition to linearly stable dynamics.
However for inhibitory networks it is known that even linearly stable networks
can display erratic dynamics (resembling chaos) due to finite amplitude 
perturbations~\cite{Zillmer2006,Timme2009ChaosBalance,MonteforteBalanced2012,angulo2014}.
This suggests that the usual linear stability analysis, corresponding to the estimation
of the maximal Lyapunov exponent~\citep{BenettinLyapunov1980}, is unable to distinguish 
between regular and irregular evolution, at least for the studied inhibitory
networks~\cite{angulo2014}.}

\red{\subsection*{Physiological relevance for biological networks under different experimental conditions}}

The analysis here reported has been inspired by the experiment performed by Carrillo {\it et al.} \cite{carrillo2008encoding}. In that experiment the authors considered a striatal network {\it in vitro}, which displays sporadic and asynchronous activity under control conditions.
To induce spatio-temporal patterned activity they perfused the 
slice preparation with N-methyl-D-aspartate (NMDA). Since it is known that
NMDA administration brings about an excitatory tonic drive with recurrent bursting activity~\cite{vergara2003,grillner2005}. The crucial role of the synaptic 
inhibition in shaping the patterned activity in striatal dynamics is also demonstrated in~\cite{carrillo2008encoding}, by applying the $GABA_a$ receptor antagonist bicuculline 
to effectively decrease the inhibitory synaptic effect.

\red{
In our simple model, ionic channels and NMDA-receptors are not modeled; nevertheless it is possible to partly} recreate the effect of NMDA administration
by increasing the excitability of the cells in the network, and the effect of the bicuculline as an effective decrease in the synaptic strength. We then verify {\it at posteriori} whether these assumptions lead to results similar to those reported in ~\cite{carrillo2008encoding}. 

In our model the single cell excitability is controlled by the parameter $I_i$. The computational experiment consists in setting the system in a low firing regime corresponding to the control conditions with $I^{(c)}= \{I_i^{(c)} \} \in [-53, \, -49.5]$ mV and in enhancing, after 20 seconds, the system excitability to the range of values $I^{(e)} = \{ I^{(e)}_i \} \in [-60, \, -45]$ mv, for another 20 seconds. This latter stage of the numerical experiment corresponds to the NMDA bath in the brain slice experiment. The process is repeated several times and the resulting raster plot is coarse grained as
explained in Methods (sub-section {\it Synchronized Event Transition Matrix}). 

\begin{figure*}
\centering
\includegraphics[width=0.25\textwidth]{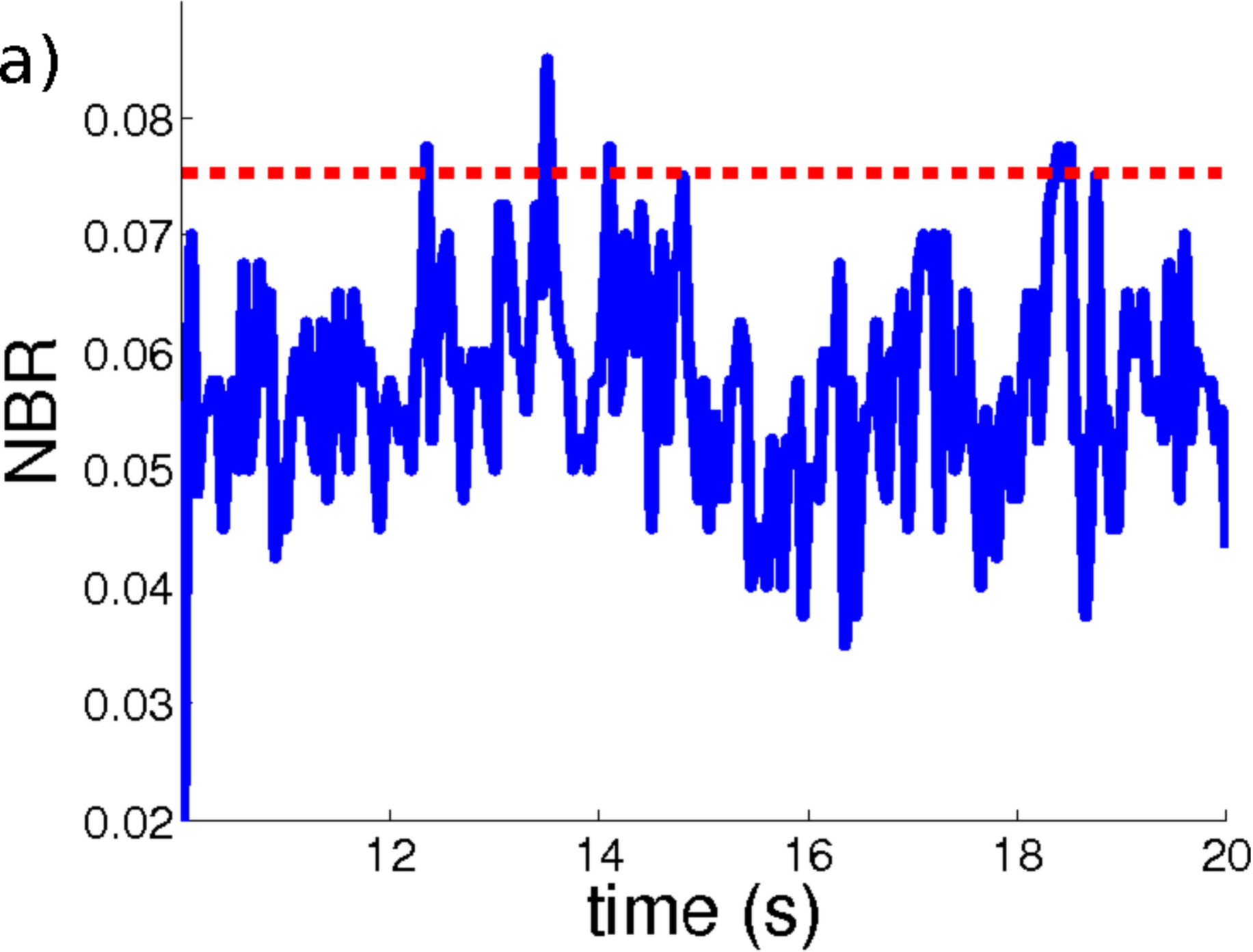}
\hspace{5mm}
\includegraphics[width=0.25\textwidth]{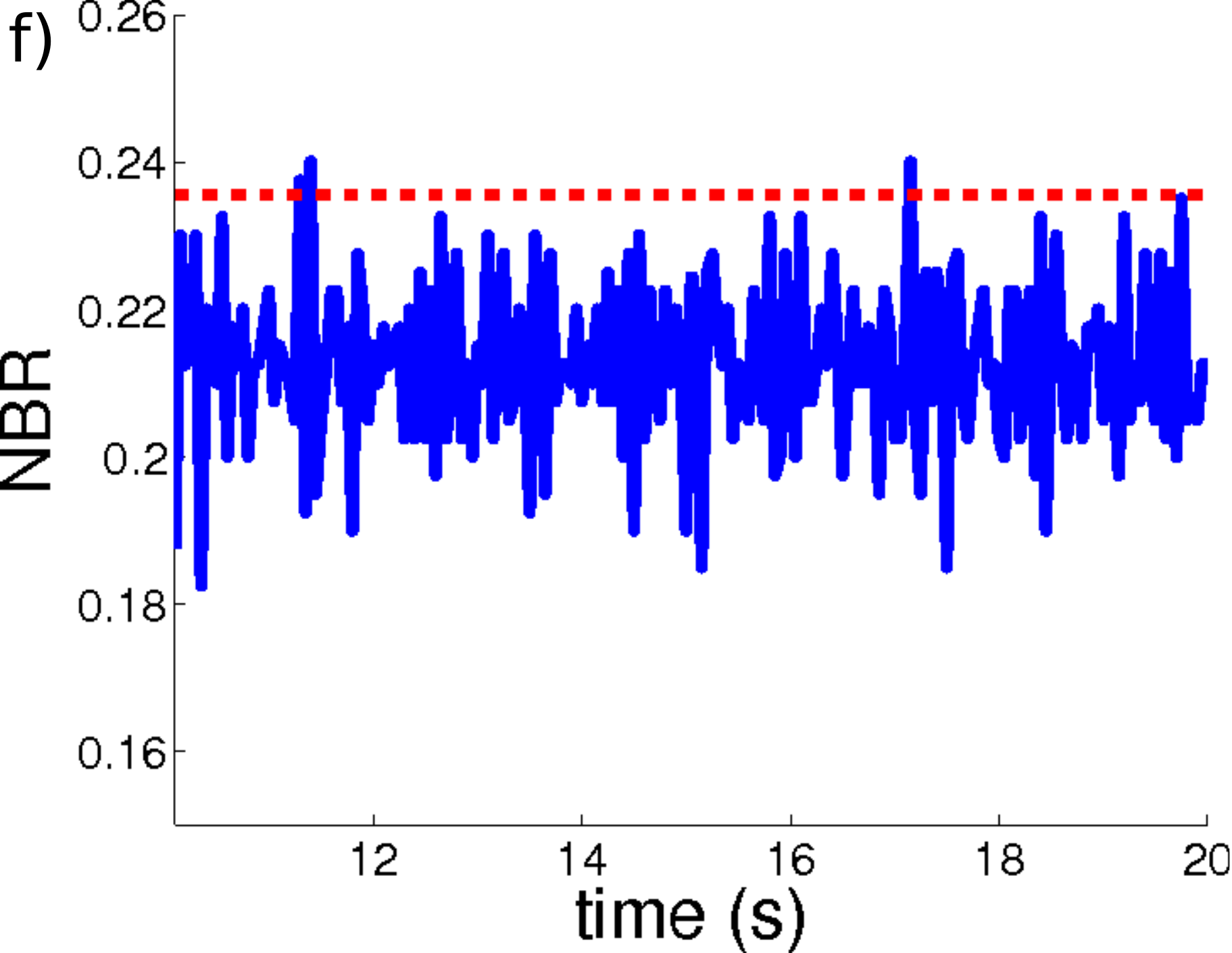}\\
\includegraphics[width=0.25\textwidth]{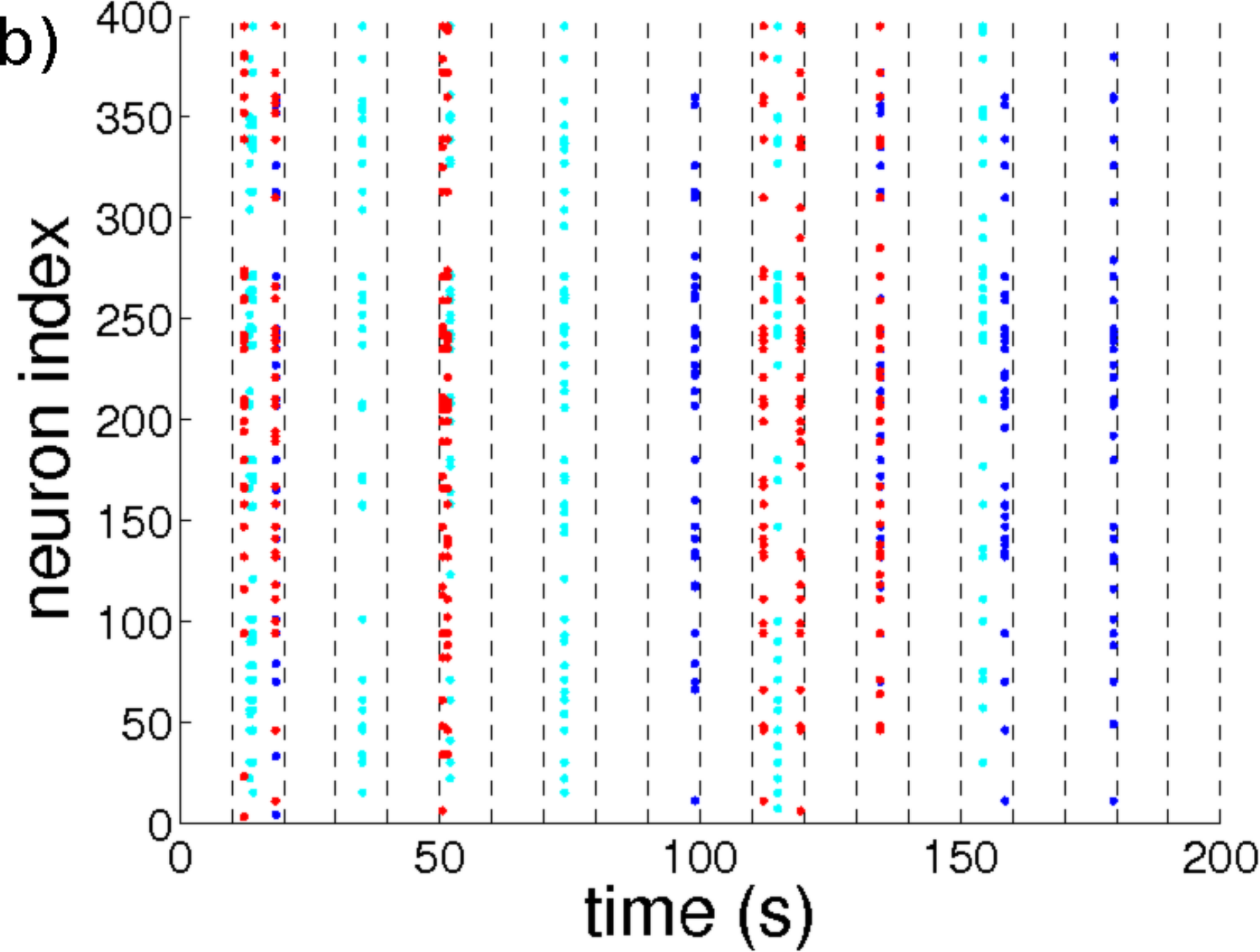}
\hspace{5mm}
\includegraphics[width=0.25\textwidth]{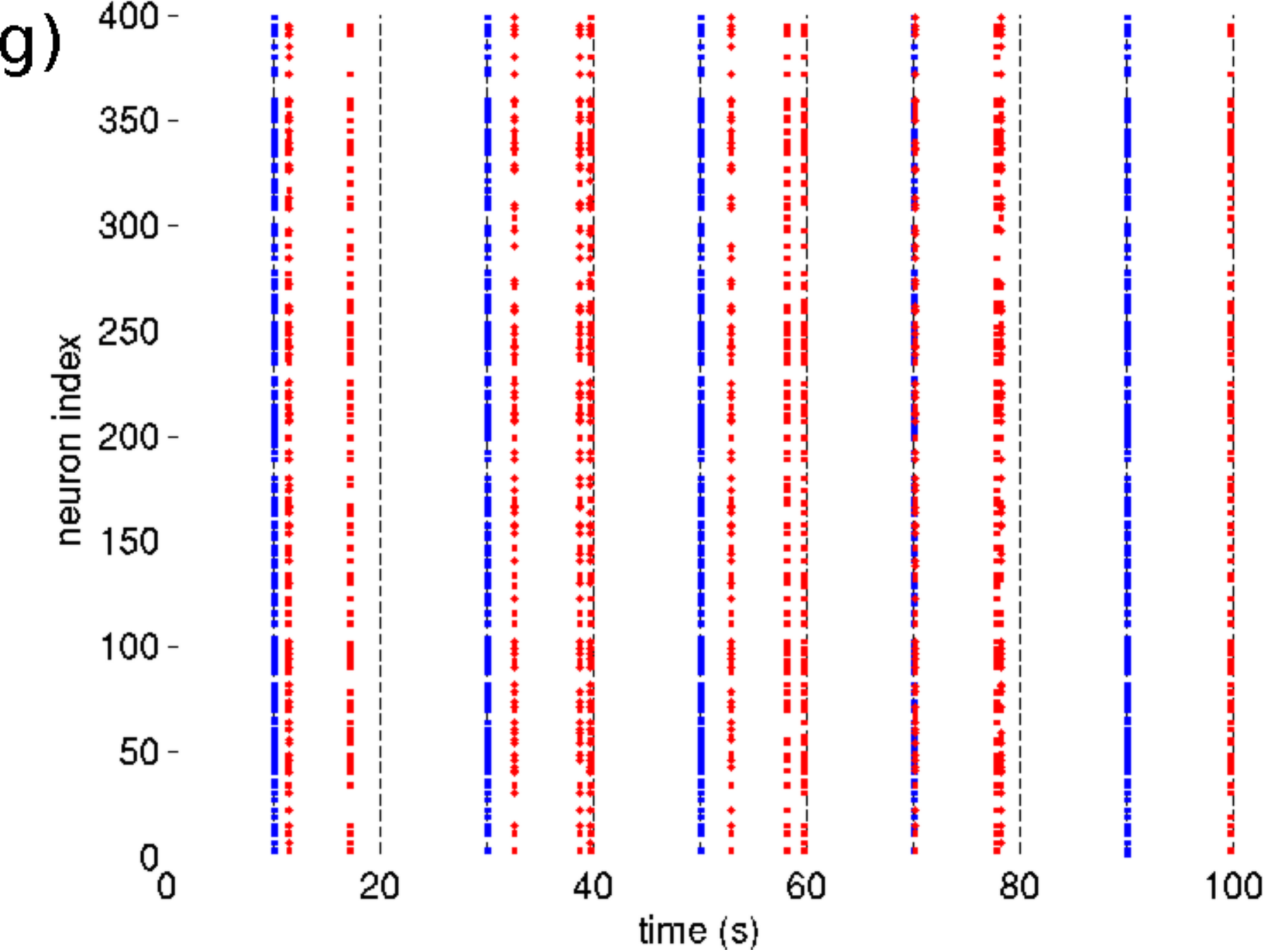}\\
\includegraphics[width=0.25\textwidth]{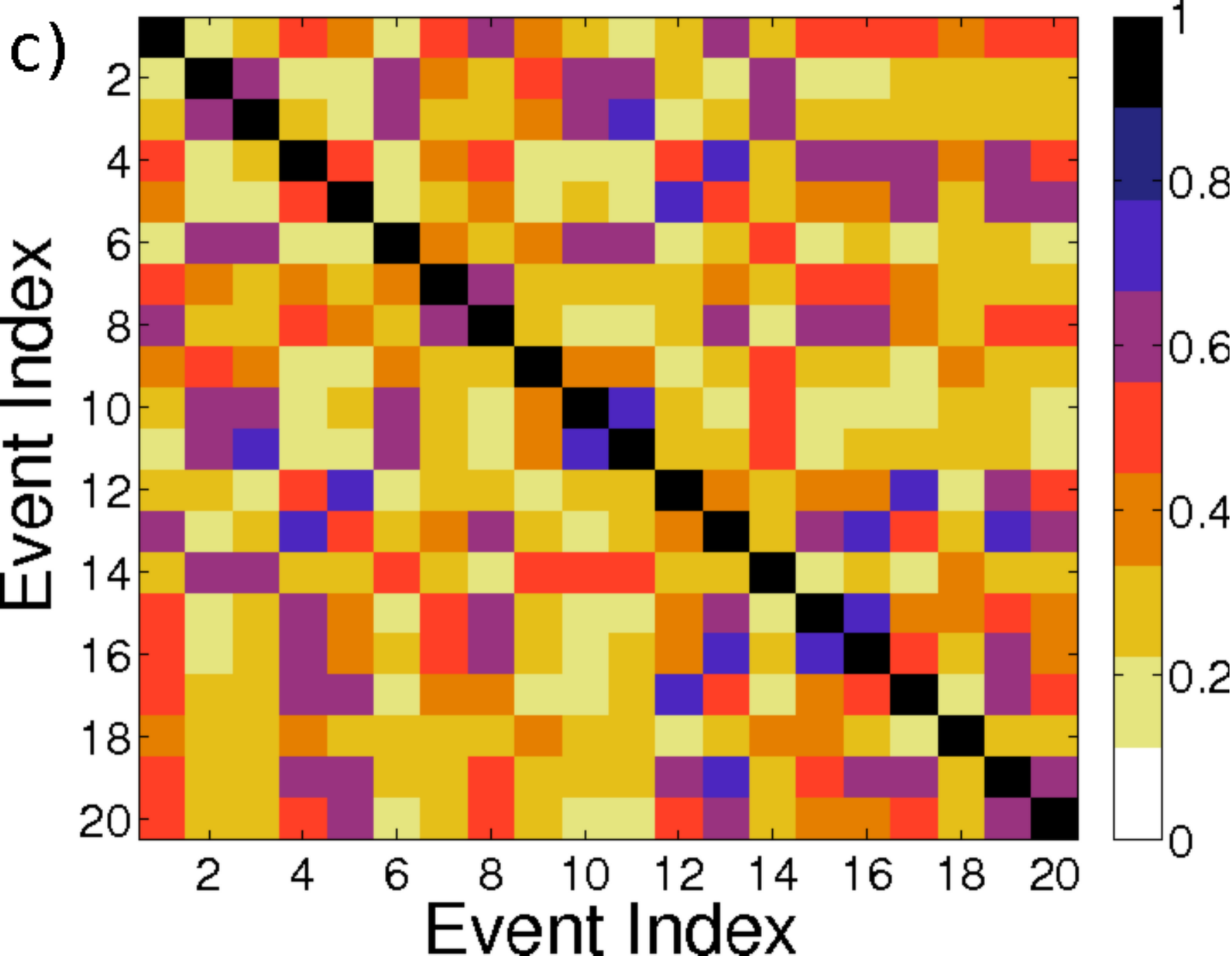}
\hspace{5mm}
\includegraphics[width=0.25\textwidth]{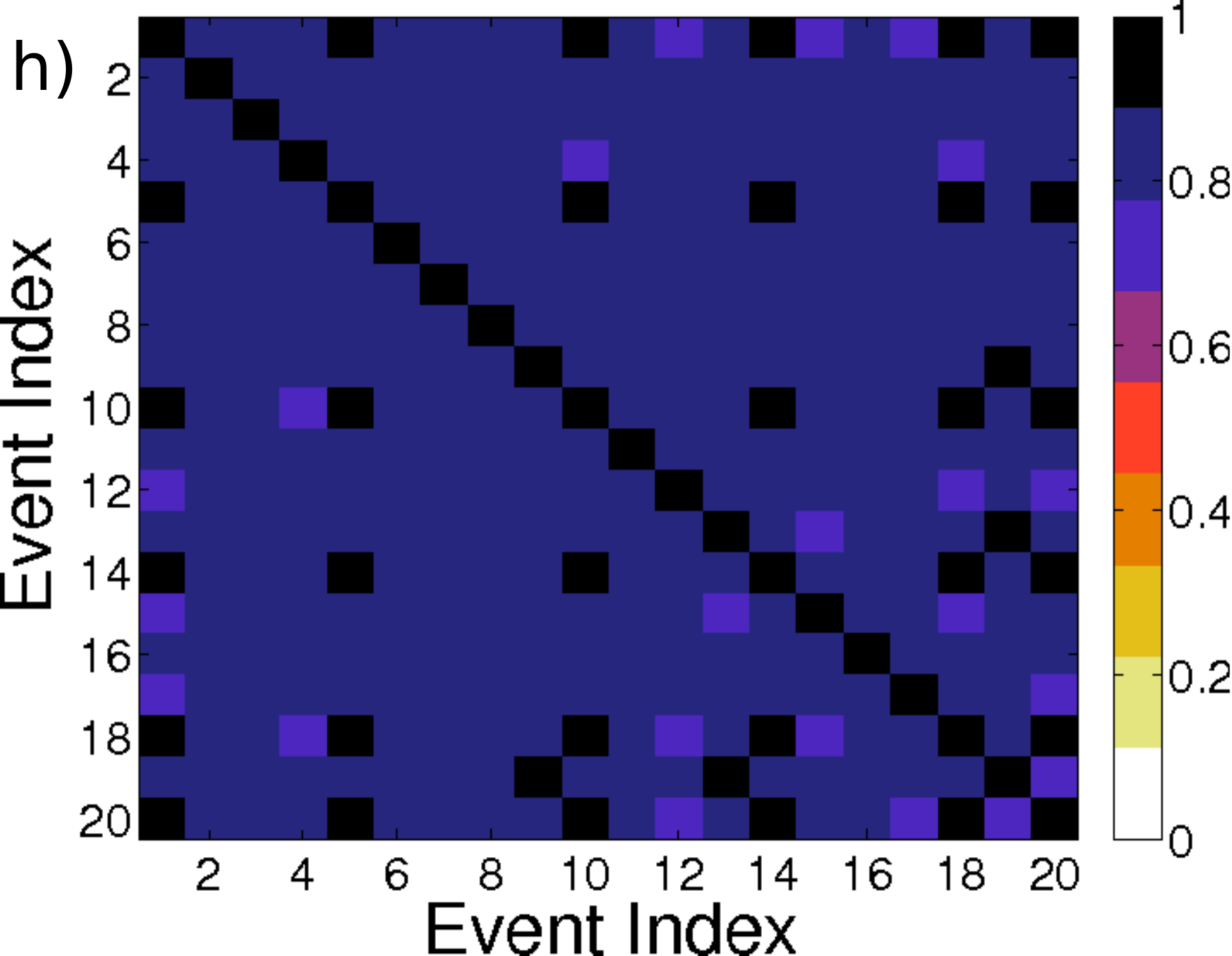}\\
\includegraphics[width=0.25\textwidth]{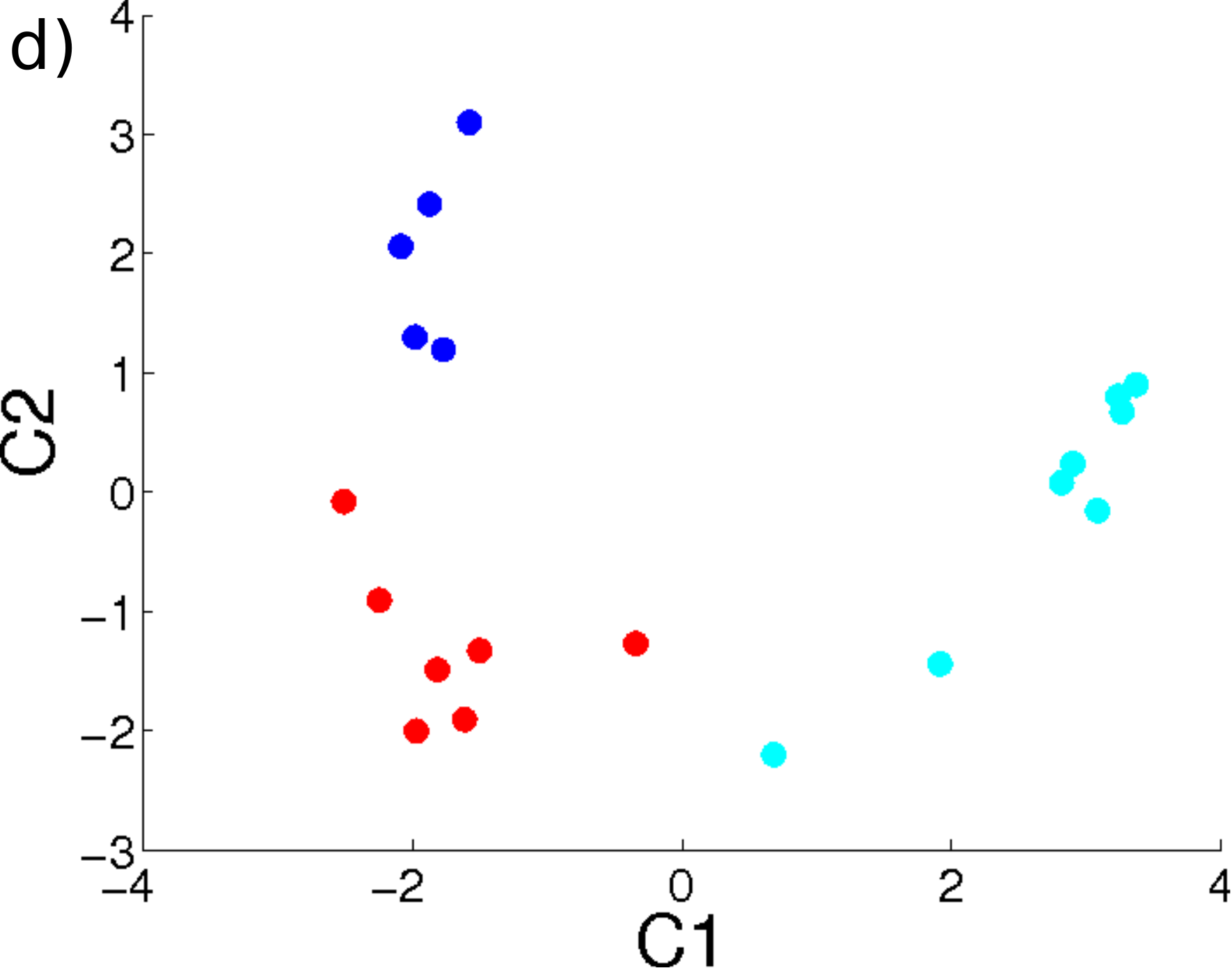}
\hspace{5mm}
\includegraphics[width=0.25\textwidth]{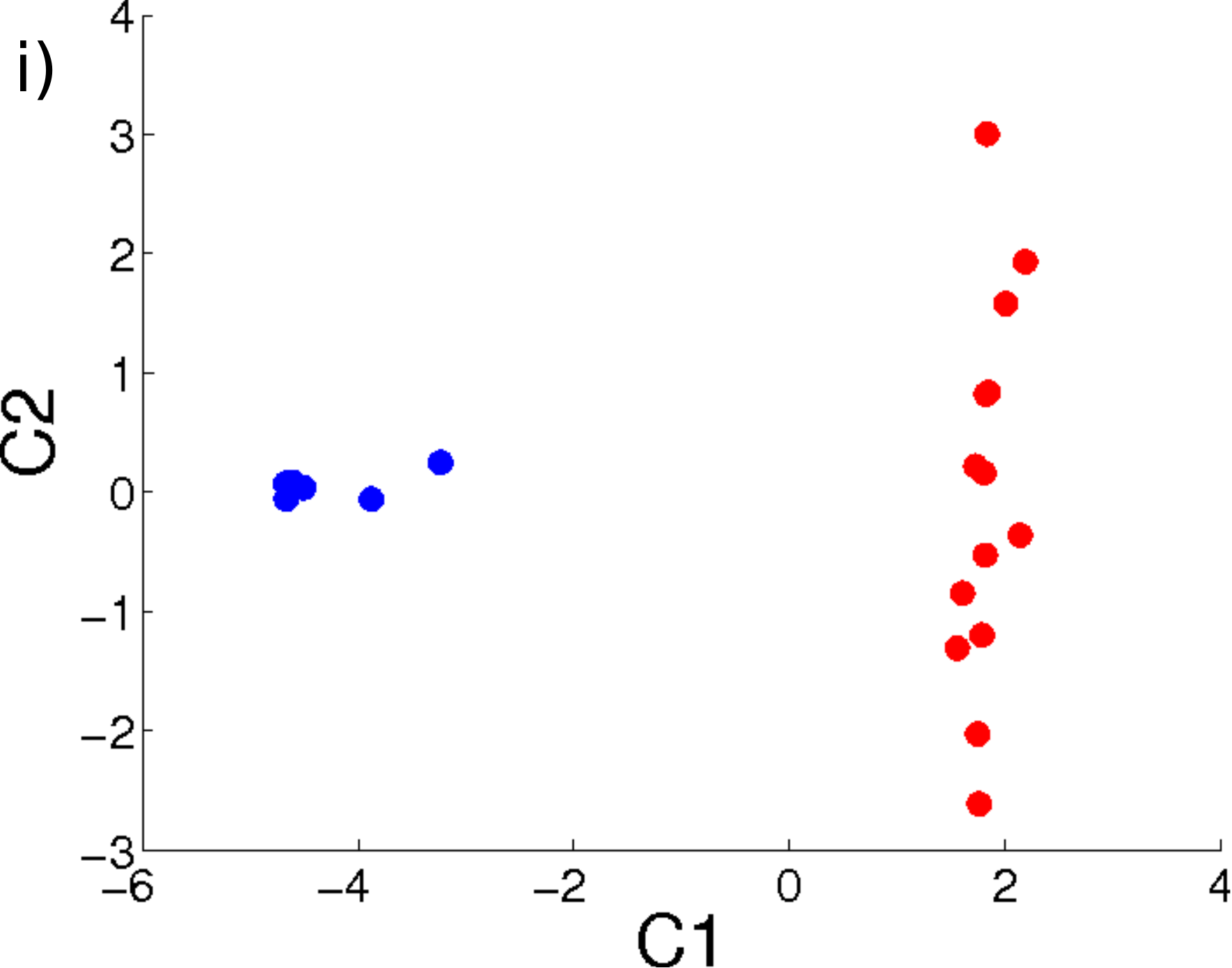}\\
\includegraphics[width=0.25\textwidth]{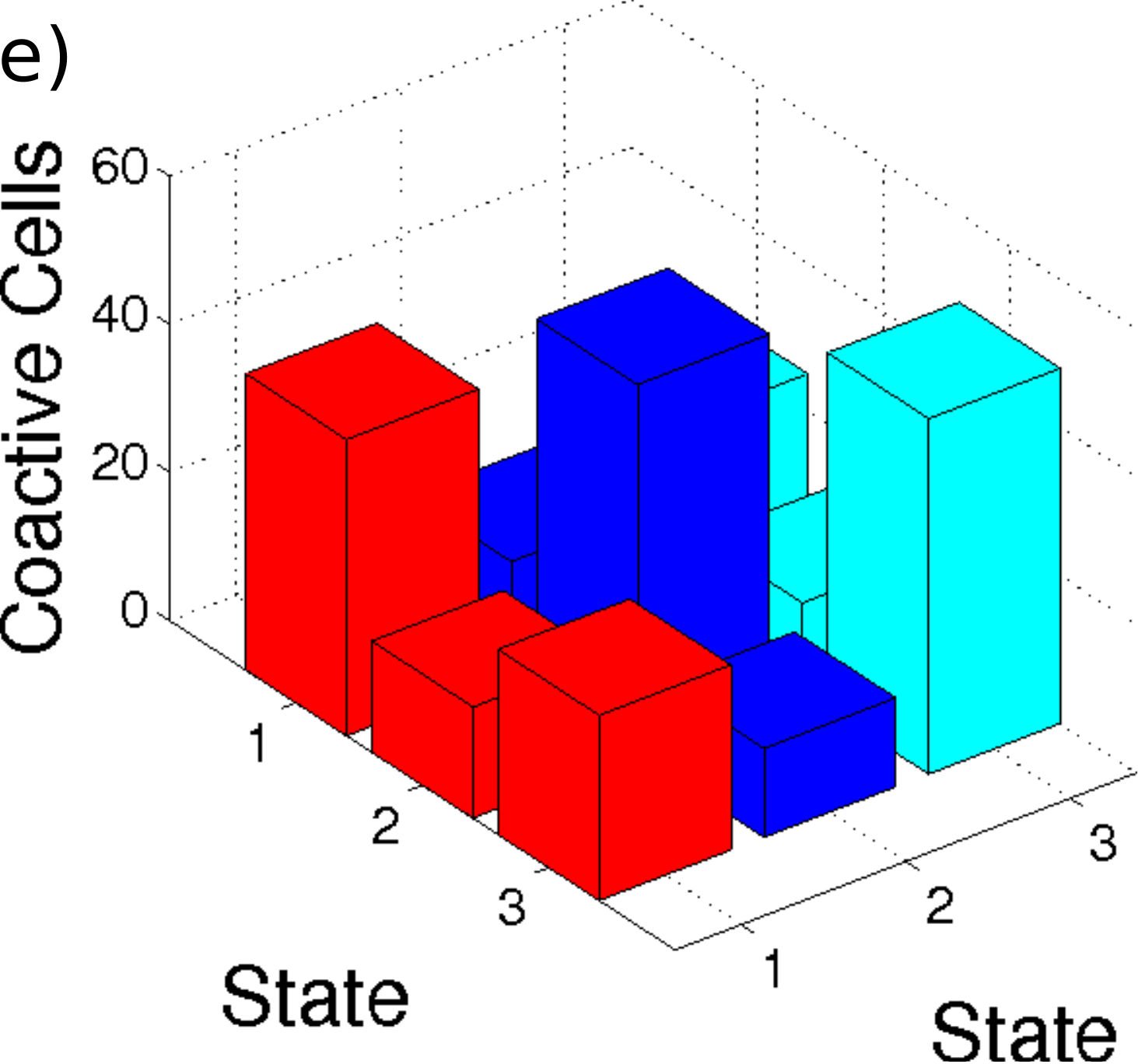}
\hspace{5mm}
\includegraphics[width=0.25\textwidth]{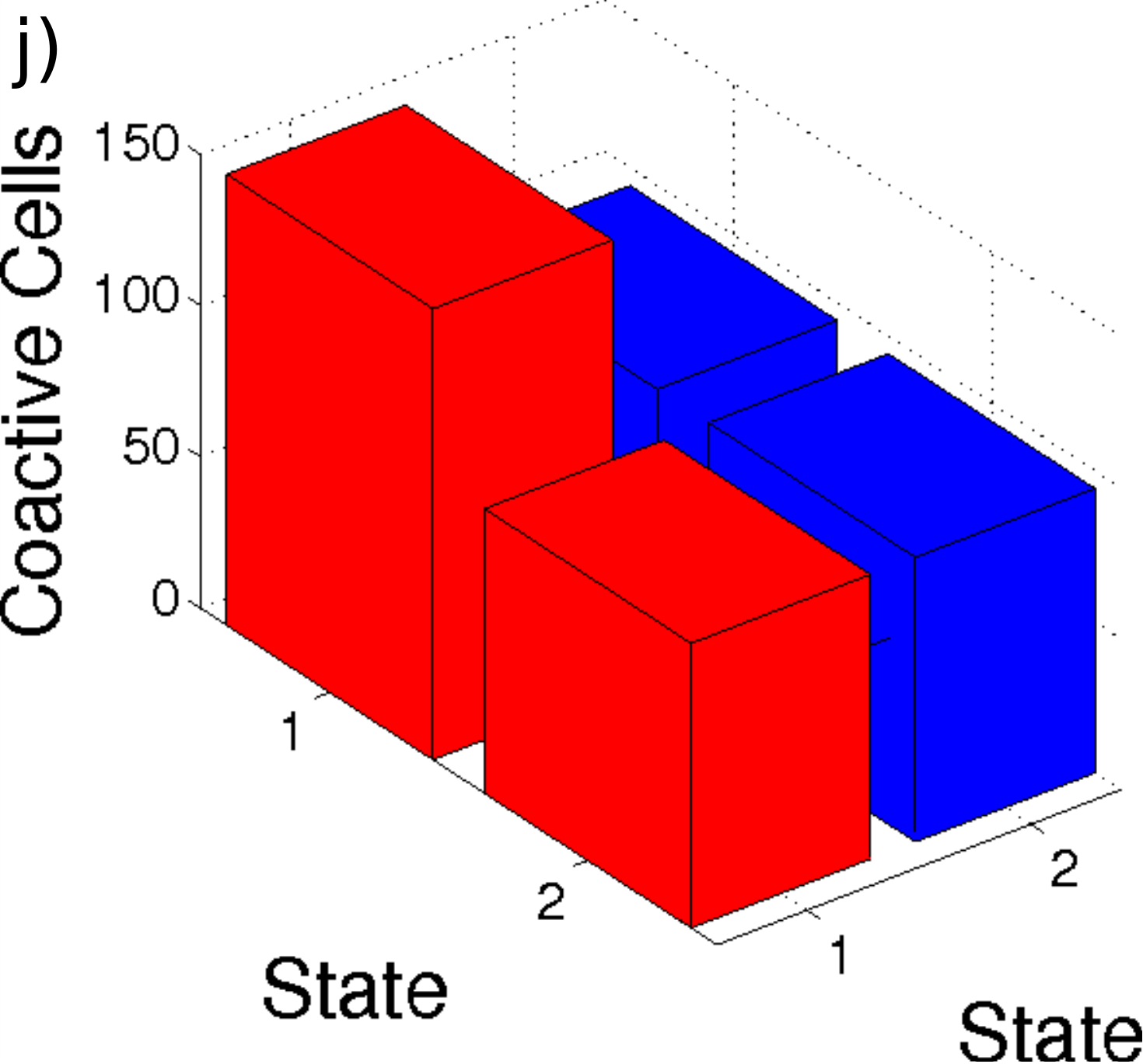}
\caption{{\bf Response of the network to an increase in the excitability.} a,f) Network Bursting Rate, and the threshold defined for considering a synchronized event. b,g) Neurons involved in the synchronized events, vertical lines denoted the switching times between the excited $I^{(e)}$ and 
control $I^{(c)}$ inputs. Colors in the raster indicates the group assigned to the synchronous event using an optimal community partition algorithm. c,h) Synchronized Event Transition Matrix, calculated with a window $T_W = 50$ ms and number of events $S_s = 20$. d,i) Projection of the synchronized 
events in the 2D space \red{spanned by the first two principal components 
associated to the covariance matrix of the vectors $W_s$}. 
e,j) Number of coactive cells in each state. The diagonal elements of the bar graph represents the total number of neurons contributing to one state. Panels (a-e) correspond to $g=8$, while panels (f-j) depict the case $g=1$. In both cases the system is recorded during the time span required to identify $S_s = 20$. Remaining parameters as in Fig. \ref{fig:PonziBenchmark}.}
\label{fig:carrillo}
\end{figure*}

From the coarse grained version of the raster plot, we calculate the {\it Network Bursting Rate} (NBR) as the fraction of neurons participating in a burst event in a certain time window. Whenever the instantaneous NBR is larger than the average NBR plus two standard deviations, this is identified as a {\it synchronized bursting event} (as shown in Fig.~\ref{fig:carrillo}(a) and (f)). In Fig.~\ref{fig:carrillo}(b) we plot all the neurons participating in a series of $S_s=20$ synchronized bursting events. 
Here the switching times between control conditions and the regimes of increased excitability are marked by vertical dashed lines. \red{Due to the choice of the parameters, the synchronized events occur only during the time intervals during which the network is in the enhanced excitability regime}. 
\red{Each synchronized event is encoded in a binary $N$ dimensional vector $W_s(t)$ with 1 (0) entries indicating that the corresponding neuron was 
active (inactive) during such event. We then measure the similarity among all the events in terms of the {\it Synchronized Event Transition Matrix} (SETM) shown
in Fig.~\ref{fig:carrillo}(c). The SETM is the normalized scalar product of all pairs of vectors $W_s$ registered in a given time interval (for more details see Methods).} 
Furthermore, using the SETM we divide the synchronized events into clusters according to an optimal clustering algorithm~\cite{newman2010networks} (see Methods). 
In the present case we have identified 3 distinct {\it states} (clusters), if we project the vectors $W_s$, characterizing each single synchronized event, on the two dimensional space spanned by the first two principal components $(C1,C2)$,  we observe a clear division among the 3 states (see Fig.~\ref{fig:carrillo}(d)). It is now important to understand whether the cells firing during the events classified as a state are the same or not. We observe that the groups of neurons recruited for each synchronized event, corresponding to a certain state, largely overlap, while the number of neurons participating to different states is
limited. As shown in Fig.~\ref{fig:carrillo}(e), the number of neurons participating to the events associated to a certain state is of the order of 40-50, while the coactive neurons (those participating in more than one state) ranges from 12 to 25. Furthermore, we have a core of 9 neurons which are firing in all states. Thus we can safely identify a distinct assembly of neurons active for each state.

As shown in  Fig.~\ref{fig:carrillo}(c), we observe, in analogy to~\cite{carrillo2008encoding}, that the system alternates its activity among the previously 
identified cell assemblies. \red{In particular, we have estimated the
transition probabilities from one state to any of the three identified states. We observe that the probability to remain in state 2 or to arrive to this state from state 1 or 3 is quite high,
ranging between 38 and 50 \%, therefore this is the most visited state. The probability that two successive events are states of type 1 or 2 is also reasonably high ranging from $\simeq 29-38 \%$ as well as the probability that from state 1 one goes to 2 or viceversa ($\simeq 38-43 \%$). Therefore the synchronized events are mostly of type 1 and 2, while the state 3 is the less {\it attractive}, since the probability of arrving to this state from the other ones or to remain on it once reached, 
are between 25 - 29 \%.} If we repeat the same experiment after a long simulation interval $t \simeq 200$ s we find that the dynamics can be always described in terms of small number of states (3-4), however the cells contributing to these states are different from the ones \red{previously} identified. This is due to 
the fact that the dynamics is in our case chaotic, as we have verified in the Supporting Information Text S1 (\textit{Linear Stability Analysis}). \red{Therefore even small differences in the initial state of the network, can have macroscopic effects on sufficiently long time scales.}
  
To check for the effect of bicuculline, the same experiment is performed again with a much
smaller synaptic coupling, namely $g=1$, the results are shown in Fig.~\ref{fig:carrillo}(f-j). The first important difference can be identified in higher NBR values with respect to the  
previous analyzed case ($g=8$) Fig.~\ref{fig:carrillo}(f). This is due to the decreased inhibitory effect, which allows most of the neurons to fire almost tonically, and therefore 
a higher number of neurons participate in the bursting events. 
In Fig.~\ref{fig:carrillo}(g) it is clearly visible a highly repetitive pattern of synchronized
activity (identified as state 2, blue symbols), this state emerges immediately after the excitability is enhanced. After this event we observe a series of bursting events,
involving a  large number of neurons (namely, 149), which have been identified as an unique
cluster (state 1, red symbols).  The system, analogously to what shown in~\cite{carrillo2008encoding}, is now locked in an unique state which is recurrently visited 
until the return to control conditions. Interestingly, synchronized events corresponding to state 1 and state 2 are highly correlated when compared with the $g=8$ case, as seen by the SETM in Fig.~\ref{fig:carrillo}(h). Despite this, it is still possible to identify both states when projected on the two dimensional space spanned by the first two principal components  (see Fig.~\ref{fig:carrillo}(i)). This high correlation can be easily explained by the fact that the neurons participating in state 2 are a subset of the neurons participating in state 1, as shown in Fig.~\ref{fig:carrillo}(j). \red{Furthermore, the analysis of the transition probabilities between states 1 and 2 reveals that starting from state 2 the system never remains in state 2, but always jumps to state 1. The probability of remaining in state 1 is really high $\simeq 64$\%. Thus we can affirm that in this case the dynamics is really dominated by the state 1.}

\red{To determine the statistical relevance of the results presented so far, we repeated the same experiment for ten different random realizations of the network,
the detailed analysis of two of these realizations is reported in Figs. S8(a-h) 
(see also Text S1).
We found that, while the number of identified states may vary from one realization to another, the persisting characteristics that distinguish the NMDA perfused scenario and the decreased inhibition one, are the variability in the SETM and the fraction of coactive cells. More precisely, on one hand the average value of the elements of the SETM is smaller for $g=8$ with respect to the $g=1$ case, namely 0.54 versus 0.84, on the other hand their standard deviation is larger, namely 0.15 versus 0.07. Thus indicating that the states observed with $g=1$ are much more correlated among them with respect to the states observable for $g=8$, which show a larger variability. The analysis of the neurons participating to the different states  
revealed that the percentage of neurons coactive in the different states passes from
51 \% at $g=8$ to 91 \% at $g=1$. Once more the reduction of inhibition leads to the emergence of states which are composed by almost the same group of active neurons, representing a dominant state. 
These results confirm that inhibition is fundamental to cell assembly dynamics.}

\red{
Altered intrastriatal signaling has been implicated in several human disorders, and in particular there is evidence for reduced GABAergic transmission following dopamine depletion~\cite{jaidar2010}, as occurs in Parkinson's disease. Our simulations thus provide a possible explanation for observations of excessive entrainment into a dominant network state in this disorder~\cite{tecuapetla2007,lopez2013}. 
}

\section{Discussion}

In summary, we have shown that \red{lateral inhibition is fundamental for
shifting} the network dynamics from a situation where a few neurons, tonically 
firing at a high rate, depress a large part of the network, to
a situation where all neurons are active and fire with similar slow rates. In particular, 
\red{if inhibition is too low, or too transient,} {\it winner takes all} mechanism is at work and
the activity of the network is mainly mean-driven. \red{By contrast, if inhibition has realistic strength and duration, almost all the neurons are on average sub-threshold
and the dynamical activity is fluctuation-driven~\cite{renart2007}.}

\red{
Therefore we can reaffirm that the MSN network is likely capable of producing slow, selective, and reproducible activity sequences as a result of lateral inhibition. The mechanism at work is akin 
to the {\it winerless competition} reported to explain
the functioning of olfactory networks in order to discriminate different odors~\cite{laurent2002olfactory}. Winnerless competition refers to a dynamical
mechanism, initially revealed in asymmetrically coupled inhibitory rate models~\cite{rabinovich2000dynamical}, displaying  a transient slow
switching evolution along a series of metastable saddles (for a recent review
on the subject see \cite{rabinovich2011robust}).
In our case,  the sequence of metastable states can be represented by the
firing activity of the cell assembly, switching over time. 
In particular, as in our analysis, slow synapses have been recognized
as a fundamental ingredient, besides asymmetric inhibitory connections, to observe
the emergence of winnerless competition in
realistic neuronal models~\cite{nowotny2007dynamical,komarov2013heteroclinic}.}

\red{
We have introduced a new metric to encompass in a single indicator
key aspects of this patterned sequential firing, and 
with the help of this metric we have
identified the values of the parameters entering in the model to best obtain these dynamics.}
 Furthermore, for the same choice of parameters 
the network is able to respond in a reproducible manner to the same stimulus, presented
at different times, while presenting complex computational capability by responding
to constant stimuli with an evolution in a high dimensional space~\cite{ostojic2014two}.

\red{
Our analysis has revelealed that the IPSP/IPSC duration is crucial in order to observe
bursting dynamics at the single cell level as well as structured assembly dynamics
at the population level. A reduction of the synaptic time has been observed
in symptomatic HD mice~\cite{cummings2010}, in our model this reduction leads
single neurons towards a Poissonian behaviour and to a reduced level of
correlation/anticorrelation among neural assemblies, in agreement with experimental results
reported for mouse models of HD~\cite{miller2008dysregulated}.}

\red{
In summary, we have been able to reproduce general experimental features 
of MSN networks in brain slices~\cite{carrillo2008encoding}. In particular, we have observed, 
as in the
experiment, a structured activity alternating among a small number of 
distinct cell assemblies. Furthermore, we have reproduced the dynamical effects
induced by decreasing the inhibitory coupling: the drastic reduction of the inhibition 
leads to the emergence of a dominant highly correlated neuronal assembly.  
This may help account for the dynamics of Parkinsonian striatal microcircuits,
where dopamine deprivation impairs the inhibitory feedback transmission 
among MSNs \cite{jaidar2010,lopez2013}. 
Network models such as the one presented here offer a path towards understanding just how pathologies that affect single neurons lead to aberrant network activity patterns, as seen in Parkinson's and Huntington's diseases, and this is an exciting direction for future research.}

\section{Methods}

\subsection*{The model}

We considered a network of $N$ LIF inhibitory neurons 
coupled via $\alpha$ pulses, which can be represented via the following set of $3N$ equations.
\begin{subequations}
\label{eq:diluted}
\begin{align}
\label{eq:dotV}
\dot{v}_i(t)&= a_i - v(t) - g E_i(t) \\
\label{eq:dotE}
\dot{E}_i(t)&= P_i(t)-\alpha E_i(t) \qquad \qquad \qquad i = 1,\dots,N \\
\label{eq:dotP}
\dot{P}_i(t)&= -\alpha P_i(t) + \frac{\alpha^2}{K} \sum_{n\mid t_n < t}\mathcal{C}_{i,j}\delta (t-t_n) \, .
\end{align}
\end{subequations}
In this model $v_i$ is the membrane potential of neuron $i$, $K$ denotes the number of pre-synaptic connections, $g > 0$ is the strength of the synapses, the variable $E_i$ accounts for the past history of all the previous recurrent post-synaptic potentials (PSP) that arrived to the neuron $i$ at times $t_n$, and $P_i$ is an auxiliary variable. $a_i$ is the external excitation and $\alpha$ is inversely proportional to the decaying time of the PSP. The inhibition is introduced via the minus sign in 
front of the synaptic strength in Eq.~(\ref{eq:dotV}). The matrix $\mathcal{C}_{i,j}$ is the connectivity matrix where the entry $i,j$ is equal to 1 if there exists a synaptic connection from neuron $j$ to neuron $i$. When the membrane potential of the $q$-th neuron arrives to the threshold $v_{th}=1$, it is reset to the value $v_r = 0$ and the cell emits \red{an $\alpha$-pulse $p_\alpha(t) = \alpha^2 t \exp{(-\alpha t)}$}  which is instantaneously transmitted to all its post-synaptic neurons. 
\red{The $\alpha$-pulses are normalized to one, therefore the area of the transmitted PSPs is
conserved by varying the parameter $\alpha$.}

The equations \eqref{eq:dotV} to \eqref{eq:dotP} can be exactly integrated from the time $t=t_n$, just after the deliver of the $n$-th pulse, to time $t=t_{n+1}$ corresponding to the emission of the $(n+1)$-th spike, thus obtaining an {\it event driven map} \cite{Zillmer2007,Olmi2010Oscillations} which reads as
\begin{subequations}
\label{eq:map}
\begin{align}
\label{eq:E_map}
E_i(n+1)&=E_i(n) {\rm e}^{-\alpha \tau(n)}+P_i(n)\tau(n) 
{\rm e}^{-\alpha \tau(n)} \\
\label{qq}
P_i(n+1)&=P_i(n) {\rm e}^{-\alpha \tau(n)}+\mathcal{C}_{iq} \frac{\alpha^2}{K}
\\
\label{V_map}
v_{i}(n+1)&=v_i(n){\rm e}^{-\tau(n)}+a(1-{\rm e}^{-\tau(n)})- g H_i(n) \, ,
\end{align}
\end{subequations}
where $\tau(n)= t_{n+1}-t_n$ is the inter-spike interval associated
with two successive neuronal firing in the network, 
which can be determined by solving the transcendental equation
\begin{equation}
\label{eq:tau_implicit}
\tau(n)=\ln\left[\frac{a-v_q(n)}{a-g H_q(n)-1}\right] \ ,
\end{equation}
here $q$ identifies the neuron which will fire at time $t_{n+1}$ by reaching the
threshold value $v_q(n+1) = 1$.

The explicit expression for $H_i(n)$ appearing in equations (\ref{V_map}) and (\ref{eq:tau_implicit}) is
\begin{eqnarray}
\label{eq:H_i}
\nonumber H_i(n) &=& \frac{{\rm e}^{-\tau(n)} - {\rm e}^{-\alpha\tau(n)}}{\alpha-1} \left(E_i(n)+\frac{P_i(n)}{\alpha-1} \right) \\
     & & -\frac{\tau(n) {\rm e}^{-\alpha\tau(n)}}{\alpha-1} P_i(n) \, .
\end{eqnarray}

The model is now rewritten as a discrete-time map with $3 N -1$ degrees of freedom, since one degree of freedom $v_q(n+1) =1$, is lost due
to the event driven procedure, which corresponds to perform a Poincar\'e section at any time a neuron fires.

The model so far introduced contains only adimensional units,
however, the evolution equation for the membrane potential (\ref{eq:diluted})
can be easily re-expressed in terms of dimensional variables as follows
\begin{equation}
\label{eq:dim}
\tau_m \dot{V}_{i}(\tilde t)= I_i - {V}_{j}(\tilde t) - 
\tau_m {G} {\tilde E}_i( \tilde t) \quad i=1,\cdots, N
\\ \quad ;
\end{equation}
where we have chosen $\tau_m = 10$ ms as the membrane time constant in agreement
with the values reported in literature for MSNs in the up state in mice~\cite{plenz1998up,klapstein2001electrophysiological, planert2013membrane}, ${I_i}$ represents the neural excitability and the external
stimulations, which takes in account the cortico-thalamic
inputs received by the striatal network. 
Furthermore, ${\tilde t} = t \cdot \tau_m$,
the field ${\tilde E}_i = E_i / \tau_m$ has the dimensionality of a frequency 
and ${G}$ of a voltage. The currents $\{I_i\}$ have also the dimensionality
of a voltage, since they include the membrane resistance.

For the other parameters/variables the transformation to physical units is simply given by
\begin{eqnarray}
{V}_{i} &=& {V}_r + ({V}_{th} - {V}_{r}) v_i\\
{I}_i &=& {V}_r + ({V}_{th} - {V}_{r}) a_i\\
{G} &=& ({V}_{th} - {V}_{r}) g 
\end{eqnarray}
where ${V}_{r}= -60$ mV, ${ V}_{th}=-50$ mV. The isolated $i$-th LIF neuron is supra-threshold
whenever $I_i > {V}_{th}$, however due to the inhibitory coupling the effective input
is $W_i = I_i - \tau_m G {\tilde E}_i$. Therefore, the neuron is able to deliver 
a spike if $\overline{W}_i  > {V}_{th}$, in this case the firing of the neuron can 
be considered as mean-driven. However, even if $\overline{W}_i  < {V}_{th}$, the neuron
can be lead to fire from fluctuations in the effective input and the firing is in this
case fluctuation-driven. It is clear that the fluctuations $\sigma(W_i)$ are directly
proportional to the strength of the inhibitory coupling for constant external currents $I_i$.

For what concerns the PSPs the associated time constant is ${\tau_\alpha} = \tau_m/\alpha$,
and the peak amplitude is given by
\begin{equation}
\label{eq:amp}
A_{PSP} = \frac{\alpha}{K} G e^{-1} = g  \times 92 \enskip \mu V \quad ;
\end{equation}
where the last equality allows for a direct transformation from adimensional
units to dimensional ones, for the connectivity considered in this paper, 
namely $K=20$, and for $\alpha=0.5$,  which is the value mainly employed in our analysis. The experimentally measured peak amplitudes of the inhibitory PSPs for spiny projection neurons ranges from $\simeq 0.16 - 0.32$ mV~\cite{tunstall2002inhibitory} to $\simeq 1-2$ mV~\cite{plenz2003}. These values depend strongly on the measurements conditions, a renormalization of all the reported measurements
nearby the firing threshold gives for the PSP peak $\simeq 0.17 - 0.34$ mV~\cite{tepper2004gabaergic}. Therefore from Eq.~(\ref{eq:amp}) one can see that
realistic values for $A_{PSP}$ can be obtained in the range $g \in [2:10]$. For $\alpha=0.5$ one gets  $\tau_{\alpha} = 20$ ms, which
is consistent with the PSPs duration and decay values reported in literature
for inhibitory transmission among spiny neurons~\cite{tunstall2002inhibitory,koos2004comparison}

\red{Our model does not take in account
the depolarizing effects of inhibitory PSPs for $V \le E_{cl}$~\cite{plenz2003}.
The GABA neurotransmitter has a depolarizing effect
in mature projection spiny neurons, however this depolarization does
not lead to a direct excitation of the spiny neurons. 
Therefore our model can be considered as an effective model
encompassing only the polarizing effects of the PSPs for $V > E_{cl}$.
This is the reason why we have assumed that the membrane potential varies in the range $[-60:-50]$ mV, 
since $E_{cl} \simeq -60$ mV and the threshold is $\simeq -50$ mV~\cite{plenz2003}.}

In the paper we have always employed dimensional variables (for simplicity we 
neglect the tilde on the time variable), apart for the amplitude of the 
synaptic coupling, for which we have found more convenient to use the adimensional quantity $g$.

\subsection*{Characterization of the firing activity}

We define active neurons, as opposite to silent neurons, 
as cells that deliver a number of spikes larger than a certain threshold $S_\Theta = 3$ during 
the considered numerical experiments. In particular, we show in Fig. S1 of the Supporting 
Information that the value of the chosen threshold does not affect the reported results 
for $0 \le S_\Theta \le 100$.

Furthermore, the characterization of the dynamics of the active neurons is performed via the coefficient of variation $CV$, the local coefficient of variation $CV_2$ and the zero lag cross-correlation matrix of the firing rates $C(\nu_i,\nu_j)$. The coefficient of variation associated to the $i$-th neuron is then defined as the ratio: 
$$CV^{(i)} = \frac{\sigma (ISI^{(i)})}{\overline{ISI^{(i)}}} \; ;$$
where $\sigma(A)$ and $ \overline{A}$ denotes the standard deviation and mean value of the
quantity $A$. The distribution of the coefficient of variation $P(CV)$ reported in the
article refer to the values of the $CV$ associated to all the active cells in the network.
\\

Another useful measure of the spike statistics is the local coefficient of variation.
For each neuron $i$ and each spike emitted at time $t^{(i)}_{n}$ from the considered
cell the local coefficient of variation is measured as

$$CV^{(i)}_{2}(n) = \frac{|ISI^{(i)}_{n} - ISI^{(i)}_{n-1}|}{ISI^{(i)}_{n}+ISI^{(i)}_{n-1}} \;$$

where the $n$-th inter-spike interval is defined as $ISI^{(i)}_{n} = t^{(i)}_{n}-t^{(i)}_{n-1}$. 
The above quantity clearly ranges between zero and one: a zero value corresponds to
a perfectly periodic firing, while the value one is attained in the limit
$ISI^{(i)}_{n}/ISI^{(i)}_{n-1} \to 0 $ (or $ISI^{(i)}_{n}/ISI^{(i)}_{n-1} \to \infty$).
The probability distribution function $P(CV_2)$ is then computed by employing the values
of the $CV^{(i)}_{2}(n)$ for all the active cells  of the network estimated at each firing event,
examples of $P(CV_2)$ are shown in Fig. ~\ref{fig:CV1_CV2}(c).

The level of correlated activity between firing rates is measured via the cross-correlation matrix $C(\nu_i,\nu_j)$. The firing rate $\nu_i(t)$ of each neuron $i$ is calculated at regular intervals
$\Delta T = 50$ ms by counting the number of spikes emitted in a time window of 
\red{$10 \Delta T = 500$ ms},
starting from the considered time $t$. 
For each couple of neuron $i$ and $j$ the corresponding element of the 
$N \times N$ symmetric cross-correlation matrix
$C(i,j)$ is simply the Pearson correlation coefficient measured as follows
$$C(i,j) = \frac{cov(\nu_i,\nu_j)}{\sigma(\nu_i)\sigma(\nu_j)} \; ,$$
where $cov(\nu_i,\nu_j)$ is the covariance between signals $\nu_i(t)$ and $\nu_j(t)$, 
\red{which has been calculated for statistical consistency by employing 
always spike trains containing 
$10^7$ events. This corresponds to time intervals $T_E$ ranging from
50 s to 350 s (from 90 s to 390 s) for $\Delta V = 5$ mV ($\Delta V =1$ mV) 
and $g \in [0.1,12]$.}

\subsection*{State Transition Matrix (STM) and measure of dissimilarity} 

The STM is constructed by calculating the firing rates $\nu_i(t)$ of the $N$ neurons at regular time intervals $\Delta T = 50$ ms. At each time $t$ the rates are measured by counting the number of spikes emitted in a window $2 \Delta T$, 
starting at the considered time. Notice that the time resolution
here used is higher with respect to that employed for the cross-correlation matrix, since we are interested in the response of the network to a stimulus presentation evaluated on a limited time window. The firing rates can be represented as a state vector $R(t) = \{ \nu_i (t) \}$ with $i=1, \dots,N$. For an experiment of duration $T_{E}$,
we have $S = \lfloor T_E/\Delta T \rfloor$  state vectors $R(t)$ representing the network evolution ($\lfloor \cdot \rfloor$ denotes the integer part).
In order to measure the similarity of two states at different times 
$t_m = m \times \Delta T$ and $t_n = n \times \Delta T$, we have calculated the
normalized scalar product
\begin{equation}
\label{eq:NormalDotProd}
D(m,n) = \frac{R(t_m) \cdot R(t_n)}{| R(t_m) | | R(t_n) |}
\end{equation}
for all possible pairs $m,n = 1, \dots, S$ during the time experiment $T_E$.
This gives rise to a $S \times S$ matrix called the state transition matrix~\cite{schreiber2003new}. 

In the case of the numerical experiment with two inputs reported in the section {\it Results}, the obtained STM has a periodic structure of period $T_{sw}$ with high correlated blocks followed by low correlated ones (see Figs. S6(b) and (e) for the complete STM). 
In Fig \ref{fig:SequantialSwitching} (b) is reported
a coarse grained version of the entire STM obtained by \red{taking a $4T_{sw} \times 4T_{sw}$ block from the STM, where the time origin corresponds to the onset
of one of the two stimuli. The block is then averaged over $r$ subsequent
windows of duration $4 T_w$, whose origin is shifted each time by $2 T_{sw}$. More precisely the averaged STM $\overline{D(m,n)}$ is obtained as follows:}
\begin{subequations}
\label{eq:lin1}
\begin{align}
\overline{D(m,n)} = \frac{1}{r^2} \sum_{i,j=1}^{r} & D(4i+m,4j+n) \\
\nonumber & \forall \, m,n \leq \lfloor T_{sw}/\Delta T \rfloor \quad.
\end{align}
\end{subequations} 

\red{In a similar manner,} we can define a dissimilarity metric to distinguish
between the response of the network to two different inputs.
We define a control input $I^{(c)} = \{I^{(c)}_i\} \in [V_{th}, \, V_{th}+\Delta V]$,
and we register the network state vectors $R^{c}(t)$ at $S$ regular time intervals
for a time span $T_{E}$. \red{We repeat the numerical experiment by considering
the same network realization and the same initial conditions, but with a new input $I^{(f)} 
= \{I^{(f)}_i\} $. The external inputs
$I^{(f)}_i$ coincide with the control ones, apart from a fraction $f$ 
which is randomly taken from the interval $[V_{th}, \, V_{th}+\Delta V]$.}
For the modified input we register another sequence $R^{f}(t)$ of state vectors on the
same time interval, with the same resolution. The instantaneous dissimilarity $d^f(t)$ 
between the response to the control and perturbed stimuli is defined as:
\begin{equation}
\label{eq:dissim}
d^f(t_m)  =  1-\frac{R^{c}(t_m) \cdot R^{f}(t_m)}{| R^{c}(t_m) | | R^{f}(t_m) |}
\end{equation}
its average in time is simply given by $\bar{d}^f  =  \frac{1}{S}\sum_{i = 1}^S d^f(t_i)$.
\red{We have verified that the average $\bar{d}^f$ is essentially not modified if
the instantaneous dissimilarities $d^f(t)$ are evaluated by considering
the state vectors $R^{c}(t_i)$ and $R^{f}(t_k)$ taken at different times within 
the interval $[0,t_S]$ and not at the same time as done in \eqref{eq:dissim}.}

\subsection*{Distinguishability metric $Q_d$:} 

Following~\cite{ponzi2013optimal} a metric of the ability of the network to distinguish 
between two different inputs $\Delta M_d$ can be defined in terms of the STM. In particular, 
let us consider the STM obtained for two inputs $I^{(1)}$ to $I^{(2)}$, each
presented for a time lag $T_{sw}$. In order to define $\Delta M_d$ the authors in
~\cite{ponzi2013optimal} have considered the correlations of the 
state vector $R$ taken at a \red{generic time $t_{m_0}$ with all 
the other configurations, with reference to Eq.~\eqref{eq:NormalDotProd} 
this amounts to examine the elements $D(m_0, n)$ of the STM $\forall t_n$.
By defining $M_1$ ($M_2$) as the average of $D(m_0,n)$ over all the times $t_n$
associated to the presentation of the stimulus $I^{(1)}$ ($I^{(2)}$), 
a distinguishablity metric between the two inputs can be finally defined as}
\begin{equation}
\Delta M_d = | M_1 - M_2 | \, .
\end{equation}

In order to take in account the single neuron variability and
the number of active neurons involved in the network dynamics
we have modified $\Delta M_d$ by multiplying this quantity
by the fraction of active neurons and the average coefficient of variation,
as follows
\begin{equation}
\label{eq:deltaM2}
Q_d = \Delta M_d \times n^*  \times \langle CV \rangle_N  \, .
\end{equation}
The above metric is reported in Figs.~\ref{fig:Q0andCV}(c),(d)
and Fig.~\ref{fig:CV1_CV2} (a).
 
\subsection*{Synchronized Event Transition Matrix (SETM)} 

In order to obtain a Synchronized Event Transition Matrix (SETM), we first
coarse grain the raster plot of the network. This is done by considering a series 
of windows of duration $T_W = 50$ ms covering the entire raster plot.
A bursting event is identified whenever a neuron $i$ fires 3 or more 
spikes within the considered window. To signal the burst occurrence
a dot is drawn at the beginning of the window. From this coarse grained raster plot 
we obtain the Network Bursting Rate (NBR) by measuring the fraction of neurons that 
are bursting within the considered window. When this fraction is larger or equal to 
the average NBR plus two standard deviations, a synchronized event is identified. 
Each synchronized event is encoded in the synchronous event vector $W_s(t)$, a $N$ dimensional binary vector where the $i$-th entry is 1 if the $i$-th neuron participated  in the synchronized event and zero otherwise. To measure the similarity between two synchronous events, we 
make use of the normalized scalar product between all the pairs of vectors $W_s$ obtained at the different times $t_i$ and $t_j$ in which 
a synchronized event occurred. This represents the element $i,j$ of the SETM.

\subsection*{Principal Components Analysis (PCA):} 

In the sub-section \textit{Discriminative and computational capability}, a Principal Component Analysis (PCA) 
is performed by collecting $S$ state vectors $R(t)$, measured at regular intervals $\Delta T$ for a time interval
$T_E$, then by estimating the covariance matrix $cov(\nu_i,\nu_j)$ associated
to these state vectors. Similarly, in the sub-section 
\textit{Physiological relevance for biological networks under different experimental conditions} 
the PCA is computed by
collecting the $S_s$ synchronous event vectors $W_s$, and the covariance matrix calculated from this set of vectors.

The principal components are the eigenvectors of theses matrices, ordered from the largest to the smallest eigenvalue. Each eigenvalue represents
the variance of the original data along the corresponding eigendirection. A reconstruction of the original data set can be obtained by projecting
the state vectors along a limited number of principal eigenvectors, obviously by employing the first eigenvectors will allow to have a more faithful reconstruction.

\subsection*{Clustering algorithms.} 

The \textit{k-means} algorithm is a widespread mining technique in which $N$ data
points of dimension $M$ are organized in clusters as follows. As a first step a number $k$
of clusters is defined a-priori, then from a sub-set of the data $k$ samples are chosen
randomly. From each sub-set a centroid is defined in the $M$-dimensional space. 
at a second step, the remaining data are assigned to the closest centroid according to a distance measure. After the process is completed, a new set of $k$ centroids can be defined by employing the data assigned to each cluster. The procedure is repeated until the
position of the centroids converge to their asymptotic value. \\

An unbiased way to define a partition of the data can be obtained by finding the optimal cluster division~\cite{newman2010networks}. The optimal number of clusters can be found by maximizing the following cost function, termed {\it modularity}:
\begin{equation}
\mathcal{M} = \frac{1}{A_{tot}} \sum_{ij} \left(A_{ij} - \mathcal{N}_{ij} \right) \delta (c_i,c_j),
\label{eq:mod}
\end{equation}
where, $A\equiv \{ A_{i,j} \}$ is the matrix to be clusterized, the normalization factor is $A_{tot} = \sum_{ij} A_{ij}$; $\mathcal{N}_{ij}$ accounts for the matrix element associated to the {\it null model}; $c_i$ denotes the cluster to which the $i$-th element of the matrix belongs to, and $\delta (i,j)$ is the Kronecker delta. In other terms, the sum appearing in Eq. (\ref{eq:mod}) is restricted to elements belonging to the same cluster.
In our specific study, $A$ is the {\it similarity matrix} corresponding to the SETM previously introduced. Furthermore, the elements of the matrix $\mathcal{N}$ are given by $\mathcal{N}_{ij} = \eta_i \eta_j/A_{tot}$, where $\eta_i = \sum_j A_{ij}$, these
correspond to the expected value of the similarity for two randomly chosen elements~\cite{newman2004fundamental,humphries2011spike}. 
If two elements are similar than expected by chance, this implies that
$A_{ij} > \mathcal{N}_{ij}$, and more similar they are larger is their contribution to the modularity $\mathcal{M}$. Hence they are likely to belong to the same cluster. The problem of modularity optimization is NP-hard \cite{fortunato2010community}, nevertheless some heuristic algorithms have been developed for finding local solutions based on greedy algorithms \cite{ronhovde2010local,reichardt2006Potts,blondel2008LouvineMethod,lancichinetti2009community}. In particular, we make use of the algorithm introduced for connectivity matrices in~\cite{newman2004fast,newman2004fundamental}, which can be straightforwardly extended to similarity matrices by considering the similarity between two elements, as the weight of the link between them~\cite{newman2004analysisWeighted}. The optimal partition technique is used in the sub-section \textit{Physiological relevance for biological networks under different experimental conditions}, where it is applied to the similarity matrix $\mathcal{S}_{ij} = 1 - \mathcal{E}_{ij}$ where the distance matrix $\mathcal{E}_{ij} = \frac{\Vert x^p_i - x^p_j \Vert_2}{\max (\mathcal{E})}$. Here $x^p_i$ is the vector defining the $i^{th}$ synchronized event projected in the first $p$ principal components, which accounts for the 80\% of the variance.

\acknowledgments

The authors had useful interactions with Robert Schmidt at an early stage
of the project. D.A.-G. and A.T. also acknowledge helpful discussions with 
Alain Barrat, Yehezkel Ben-Ari, Demian Battaglia, Stephen Coombes, Rosa Cossart, Diego Garlaschelli, Stefano Luccioli, Mel MachMahon, Rossana Mastrandea, Ruben Moreno-Bote, and Viktor Jirsa. This work has been partially supported by the European Commission under the program ``Marie Curie Network for Initial Training", through the project N. 289146, ``Neural Engineering Transformative Technologies (NETT)"
(D.A.-G. and A.T), by the A$^\ast$MIDEX grant (No. ANR-11-IDEX-0001-02) 
funded by the French Government ``program Investissements d'Avenir'' (A.T.),
by US National Institutes of Health awards NS078435, MH101697, DA032259 and NS094375 (J.B.),
and by ``Departamento Adminsitrativo de Ciencia Tecnologia e Innovacion - Colciencias" through
the program ``Doctorados en el exterior - 2013" (D.A.-G.).

\end{document}